\documentclass[12pt]{article}
\pdfoutput=1
\usepackage{jheppub}

\usepackage{amsmath,amssymb,amsfonts}
\usepackage{graphicx}
\usepackage{color}
\usepackage{tikz}
\usepackage{dsfont}
\usepackage[normalem]{ulem}
\usepackage{booktabs}

\usepackage{multirow}
\usepackage[all]{hypcap}

\usepackage{subfigure}

\usepackage{comment}

\makeatletter\def\l@subsubsection#1#2{}
\makeatother

\def\be{\begin{eqnarray}}
\def\ee{\end{eqnarray}}

\def\makeatletter{\catcode`\@=11}
\makeatletter
\def\mathbox#1{\hbox{$\m@th#1$}}%
\def\math@ccstyles#1#2#3#4#5#6#7{{\leavevmode
      \setbox0\mathbox{#6#7}%
      \setbox2\mathbox{#4#5}%
      \dimen@ #3%
      \baselineskip\z@\lineskiplimit#1\lineskip\z@
      \vbox{\ialign{##\crcr
             \hfil \kern #2\box2 \hfil\crcr
             \noalign{\kern\dimen@}%
             \hfil\box0\hfil\crcr}}}}
\def\mathaccstyles{\math@ccstyles\maxdimen}
\def\maththroughstyles{\math@ccstyles{-\maxdimen}}
\def\unity%
 {\maththroughstyles{.45\ht0}\z@\displaystyle {\mathchar"006C}\displaystyle 1}

\begin{document}
\title{String theory and the SymTFT of 3d orthosymplectic Chern-Simons theory}

\author[a, b]{Oren Bergman,}
\emailAdd{bergman@physics.technion.ac.il}
\author[a, c]{Francesco Mignosa}
\emailAdd{francesco.mignosa02@gmail.com}

\affiliation[a]{Department of Physics, Technion, Israel Institute of Technology,\\
Haifa, 32000, Israel\\[-2mm]}
\affiliation[b]{Department of Physics, University of California,\\
San Diego, La Jolla, CA 92093, USA\\[-2mm]}
\affiliation[c]{Department of Physics, Universidad de Oviedo,\\
C/Federico Garcia Lorca 18, 33007 Oviedo,
Spain\\[-2mm]}

\abstract{We revisit the 3d ${\cal N}=5$ Chern-Simons-Matter theory with orthosymplectic gauge group and its gravity dual
from the perspective of generalized symmetries.
We derive the corresponding 4d symmetry topological field theory from the gravity dual
and relate the allowed boundary conditions to the different variants of the 3d theory.
Concentrating on a specific variant that has a discrete non-abelian global symmetry, we explain how the structure 
of this symmetry arises from brane dynamics.}

\maketitle

\vspace{1.6 cm}
\vfill

\newpage

\section{Introduction}
\label{intro}

The concept of symmetry topological field theory (SymTFT) has emerged as a powerful
organizational tool for distilling the symmetries and anomalies of quantum field theories (QFT) 
\cite{Freed:2012bs,Apruzzi:2021nmk}.
This is a generalization of the anomaly theory, a topological field theory in $M_{d+1}$ that encodes
the anomalies of the global symmetries of a given QFT in $M_d = \partial M_{d+1}$.
The SymTFT also lives in a $(d+1)$-dimensional spacetime $M_{d+1}$, which is now thought to have two
boundaries: a physical boundary where the dynamics of the QFT occur, and a topological boundary that encodes 
the global symmetries and their properties.
This defines a class of QFTs with the same local dynamics, but with possibly different global symmetries and anomalies,
referred to as a {\em relative QFT}. 
A specific {\em absolute QFT} is then defined by specifying boundary conditions
for the bulk gauge fields at the topological boundary. 
In particular, the boundary conditions determine the global symmetries of the
boundary QFT.
Depending on the boundary conditions, a given topological operator of the SymTFT will correspond in the boundary theory to either 
a symmetry operator or to an operator on which a symmetry acts. 
In the former case, this is represented by an operator brought to the topological boundary, and in the latter case, by an operator
ending at the topological boundary.
From the point of view of the $d$-dimensional boundary QFT, 
the different absolute theories are related to each other by gauging different parts of the global symmetry,
which from the point of view of the bulk SymTFT corresponds to exchanging Neumann and Dirichlet boundary conditions
for the associated gauge fields.

In fact this idea has its origins in {\em AdS/CFT}, the class of dualities relating superstring theory in $AdS_{d+1}$ to 
conformal field theories in $d$ spacetime dimensions.
Near the boundary of $AdS_{d+1}$ string theory reduces generically to a $(d+1)$-dimensional topological field theory 
of the bulk gauge fields. Moreover, the formulation of the theory in $AdS$ space 
requires the specification of boundary conditions for the bulk gauge fields, which in turn determine the 
precise version of the dual CFT. This was first shown by Witten for 4d ${\cal N}=4$ SYM theory with 
$\mathfrak{su}(N)$ gauge algebra \cite{Witten:1998wy}.
The 5d theory obtained by reducing Type IIB supergravity to $AdS_5$ is precisely the SymTFT of the 4d gauge theory,
and the set of consistent boundary conditions on the bulk gauge fields reproduces the different possible global structures
of the gauge group \cite{Witten:1998wy,Aharony:2013hda,Bergman:2022otk}.
SymTFT's were subsequently derived for  
several other examples of both holographic and geometrically engineered QFTs in various dimensions
\cite{Bergman:2020ifi,Apruzzi:2021nmk,Apruzzi:2022rei,Heckman:2022xgu,Argurio:2024oym}. 

In these string theory settings, the topological operators of the SymTFT are realized as branes wrapping various cycles
of the internal space. It has long been appreciated that branes that extend in the radial direction and end at the boundary
of $AdS$ correspond to charged operators in the boundary theory. For example strings ending at the boundary of $AdS_5$ 
describe Wilson lines in the 4d gauge theory \cite{Maldacena:1998im}.
But it has also become apparent more recently that the symmetry operators that act on these charged operators are also
described by branes, that are now transverse to the radial direction, and taken to the boundary
\cite{OBtalks,Apruzzi:2022rei,GarciaEtxebarria:2022vzq,Bergman:2022otk,Heckman:2022muc,Apruzzi:2023uma, Bah:2023ymy}.\footnote{This is most well understood
in the case of finite symmetries. However see \cite{Cvetic:2023plv} and \cite{Bergman:2024aly} for recent proposals regarding $U(1)$ symmetries.}
The boundary condition on the gauge field under which a given brane is charged determines its role in the boundary theory,
as a charged operator or as a symmetry operator.

In this paper we will add another interesting case study to the mix:
the 3d ${\cal N}=5$ supersymmetric orthosymplectic quiver gauge theory $\mathfrak{so}(2n)_{2k} \times \mathfrak{usp}(2n)_{-k}$ 
of \cite{Hosomichi:2008jb,Aharony:2008gk}.
This is the low energy effective theory that arises on the worldvolume of $N$ M2-branes probing a $\mathbb{C}^4/\hat{\mathbb{D}}_k$ 
singularity (where $\hat{\mathbb{D}}_k$ is the binary dihedral group).
At large $N$ this theory is dual to M-theory on $AdS_4\times S^7/\hat{\mathbb{D}}_k$, or alternatively to Type IIA
string theory on $AdS_4\times \mathbb{C}P^3/\mathbb{Z}_2$ \cite{Aharony:2008gk}.
We therefore expect to recover the 4d SymTFT by reducing 11d supergravity on $S^7/\hat{\mathbb{D}}_k$,
or 10d Type IIA supergravity on $\mathbb{C}P^3/\mathbb{Z}_2$. This is precisely what we will show, at least 
in the Type IIA description.

The 3d orthosymplectic theories exhibit interesting categorical symmetry structures depending on the specific global
form of the gauge group.
Some of these structures were uncovered in \cite{Beratto:2021xmn, Mekareeya:2022spm} from the perspective of the superconformal index.
In one case, the theory has a finite non-abelian global symmetry given by either the dihedral group $\mathbb{D}_8$
or the quaternion group $\mathbb{Q}_8$.
The appearance of the dihedral group as a global symmetry was previously shown in a slightly different setting 
in \cite{Gaiotto:2017yup}, and later in \cite{Bhardwaj:2022yxj, Bhardwaj:2022maz, Bartsch:2022ytj} for the purely orthogonal theory 
with gauge group $SO(2n)/\mathbb{Z}_2$.
The latter references also studied other global forms of the orthogonal gauge group and how they are related by 
gauging parts of the global symmetry.

The orthosymplectic theories that we study here generalize and extend these results, for example to include 
a case with $\mathbb{Q}_8$ global symmetry, to theories with a known holographic dual.
This will allow us to derive the SymTFT from the reduction of Type IIA supergravity on $AdS_4\times \mathbb{C}P^3/\mathbb{Z}_2$, 
and to recover the symmetry structures
from the properties of the various branes in this background.

The rest of the paper is organized as follows.

In section \ref{symTFTD8Q8} we construct the 4d symmetry theory corresponding to the global symmetries 
and anomalies of the type we are interested in, without reference to any explicit 3d QFT.
We present the topological operators of the theory, determine their link-pairings and fusion relations,
and show that one choice of boundary conditions gives rise to a non-abelian global symmetry, $\mathbb{D}_8$ or $\mathbb{Q}_8$.
In section \ref{orthosymplecticanalysis} we discuss an explicit class of 3d QFT's that have this structure,
the ${\cal N}=5$ orthosymplectic quiver theories, and we relate the different boundary conditions in the SymTFT to the
different global structures of the gauge group.
In section \ref{stringtheory} we review the Type IIA string theory dual of the 3d orthosymplectic theories, and derive the 
4d SymTFT by dimensionally reducing Type IIA supergravity on the internal space $\mathbb{C}P^3/\mathbb{Z}_2$.
Along the way we clarify a point of confusion from \cite{Aharony:2008gk}.
In section \ref{branes} we discuss the realization of the SymTFT operators in terms of branes in the Type IIA string theory background.
In particular we show how brane dynamics reproduce the non-abelian symmetries $\mathbb{D}_8$ and $\mathbb{Q}_8$.
Finally, section \ref{concl} contains our conclusions and outlook.

\section{Symmetry}
\label{symTFTD8Q8}



In a landmark paper \cite{Tachikawa:2017gyf} Tachikawa showed that anomalies of finite symmetries in QFT are related to extension classes 
of finite quotient groups.
Consider a three-dimensional theory ${\cal T}$ with a discrete, anomaly free, 0-form symmetry group $\Gamma$ that has a normal abelian subgroup $H$.
Gauging the subgroup $H$ leads to a theory ${\cal T}/H$ with a 0-form symmetry $G = \Gamma/H$ and a 1-form symmetry $H$.
The 0-form and 1-form symmetries have a mixed anomaly specified by the extension class $e \in H^2(G,H)$ associated to the exact sequence
\be
1 \rightarrow H \rightarrow \Gamma \rightarrow G= \Gamma/H \rightarrow 1 \,.
\ee
This anomaly can be expressed in terms of a 4d anomaly theory given by
\be
\label{SimpleAnomalyAction}
S_4^{anomaly}[A_2,A_1] = \int_{M_{4}} A_{2} \cup e(A_1) = \int_{M_{4}} A_{2} \cup A_1 \cup A_1
\ee
where $A_2$ is a flat background connection (a 2-cochain) for the 1-form symmetry $H$, and $A_1$ is a flat background connection 
(a 1-cochain) for the 0-form symmetry $G$.
Going the other way, if we start with a theory that has a finite 0-form symmetry $G$ and a finite 1-form symmetry $H$,
with a mixed anomaly given by eq. (\ref{SimpleAnomalyAction}), then gauging the 1-form symmetry leads 
to a 0-form symmetry $\Gamma$ which is an extension of $G$ by $H$ corresponding to the extension 
class $e$.\footnote{In other cases this can lead to a higher-group symmetry \cite{Tachikawa:2017gyf}.}

For example if $G=\mathbb{Z}_2$ and $H=\mathbb{Z}_2$ there are two possibilities classified by 
\be
\label{SimpleExtensionClass}
H^2(\mathbb{Z}_2,\mathbb{Z}_2) = \mathbb{Z}_2 \,.
\ee
They are the trivial extension $\Gamma = \mathbb{Z}_2 \times \mathbb{Z}_2$, and the non-trivial extension $\Gamma = \mathbb{Z}_4$.
In the non-trivial extension, the order 2 element of $G$ is ``fractionalized" to an order 4 element.
The case we are interested in is $G = \mathbb{Z}_2 \oplus \mathbb{Z}_2$ and $H = \mathbb{Z}_2$.
In this case there are eight extensions classified by
\be
H^2(\mathbb{Z}_2\oplus \mathbb{Z}_{2},\mathbb{Z}_2) = \mathbb{Z}_{2} \oplus \mathbb{Z}_{2}  \oplus \mathbb{Z}_{2} \,.
\ee
Denoting an element of this group by $(\alpha,\beta,\gamma)$, the trivial extension $\Gamma = \mathbb{Z}_2 \oplus \mathbb{Z}_2 \oplus \mathbb{Z}_2$
corresponds to $(0,0,0)$. The remaining non-trivial extensions are shown in Table~\ref{Extensions}.
Note that there are three ways to get $\Gamma = \mathbb{Z}_4\oplus \mathbb{Z}_2$, and three ways to get $\mathbb{D}_8$, which also has a single order 4 element.
These are associated to which of the three order 2 elements of $G$ get fractionalized to the order 4 element.
The quaternion group $\mathbb{Q}_8$ has three order 4 elements. All three order 2 elements of $G$ are fractionalized in this case.
\begin{table}[h!]
\centering
\begin{tabular}{ |c|c|} 
\hline
$(\alpha,\beta,\gamma)$ & $\Gamma$ \\ 
 \hline
 $(0,0,0)$ & $\mathbb{Z}_2  \oplus \mathbb{Z}_2 \oplus \mathbb{Z}_2$ \\
 $(1,0,0),(0,1,0),(1,1,0)$ & $\mathbb{Z}_4\oplus \mathbb{Z}_2$ \\
 $(0,0,1),(1,0,1),(0,1,1)$ & $\mathbb{D}_8$ \\
 $(1,1,1)$ & $\mathbb{Q}_8$\\
 \hline
\end{tabular}
\caption{The possible extensions of $\mathbb{Z}_2 \oplus \mathbb{Z}_2 \oplus \mathbb{Z}_2$.
$\mathbb{D}_8$ is the dihedral group of order eight, and $\mathbb{Q}_8$ is the quaternion group.
Their properties are summarized in Appendix~\ref{Groups}.}
\label{Extensions}
\end{table}
The anomaly theory is given by 
\be
\label{GeneralAnomalyAction}
S_4^{anomaly}[A_{2}^B,A_1^M,A_1^C] &=& i\pi \int_{M_{4}} A_{2}^B \cup \left(\alpha A_1^{M} \cup A_1^{M} + \beta A_1^{C} \cup A_1^{C}
+ \gamma A_1^{M} \cup A_1^{C}\right) \nonumber \\
&=& i \pi \int_{M_{4}} A_{2}^B \cup \left({\alpha\over 2} \delta A_1^{M}  + {\beta\over 2} \delta A_1^{C} 
+ \gamma A_1^{M} \cup A_1^{C}\right) \,,
\ee
where $A_2^B$ is the background field for the $\mathbb{Z}_2$ 1-form symmetry, and $A_1^M, A_1^C$ are the background fields for
the two $\mathbb{Z}_2$ factors in the 0-form symmetry.\footnote{The second expression for the anomaly action in eq. (\ref{GeneralAnomalyAction})
is understood as follows.
The first two terms can be expressed in terms of the extension class in eq. (\ref{SimpleExtensionClass}), as in eq. (\ref{SimpleAnomalyAction}).
This measures the obstruction to lifting a $\mathbb{Z}_2$ bundle to a $\mathbb{Z}_4$ bundle.
Given a $\mathbb{Z}_2$ bundle described by a $\mathbb{Z}_2$ 1-cochain $A_1$ with $\delta A_1 = 0$, namely a 1-cocycle,
its lift $\tilde{A}_1$ to $\mathbb{Z}_4$ satisfies $\delta\tilde{A}_1 = 2 e(A_1)$.}

\subsection{The symmetry theory and symmetry operators}

The SymTFT generalizing the anomaly theory in eq. (\ref{GeneralAnomalyAction})
is obtained by adding $BF$ type terms for the three gauge fields:
\be \label{SymTFT1}
S^{\text{sym}}_4 =  S^{\text{anomaly}}_4+ i \pi \int_{M_4} \left(A_2^B\delta A_1^B + A_{2}^{M} \delta A_{1}^{M} 
+ A_{2}^{C} \delta A_{1}^{C}\right) \,,
\ee
where $A_1^B$, $A_2^M$, and $A_2^C$ are three additional cochains, that can be thought of as sources for the three original cochains.
The cup product symbol has been dropped for conciseness.
From now on we will assume that $\gamma = 1$, namely that the triple anomaly defined by the cubic term in the anomaly action
(\ref{GeneralAnomalyAction}) is nontrivial.
The SymTFT is therefore given by
\be
\label{SymTFT2}
S^{\text{sym}}_4  =  i \pi \int_{M_4} \left[A_2^B \delta \left(A_1^B + {\alpha\over 2}A_1^M + {\beta\over 2}A_1^C \right)
+ A_{2}^{M} \delta A_{1}^{M} + A_{2}^{C} \delta A_{1}^{C} +  A_2^B A_1^M A_1^C \right] .
\ee
A 3d version of this SymTFT was studied in \cite{Kaidi:2023maf}.
Much of what follows in this section is a straightforward generalization of that paper.

The gauge transformations of the original fields are as usual,
\be \label{gaugetransf}\nonumber
A_2^B & \rightarrow &  A_2^B + \delta \Lambda_1^B \\ 
A_1^{M} & \rightarrow & A_1^{M} + \delta \Lambda_0^{M} \\
A_1^{C} & \rightarrow & A_1^{C} + \delta \Lambda_0^{C} \,, \nonumber
\ee 
but those of the additional fields are modified due the presence of the cubic term in the action:
\be \nonumber
A_1^B & \rightarrow & A_1^B + \delta \Lambda_0^B
- \Lambda_0^{M} A_1^{C} 
+ \Lambda_0^{C} A_1^{M} 
- \Lambda_0^{M}\delta\Lambda_0^{C}\\ \label{anomalousgtransformations}
A_2^{M} & \rightarrow & A_2^{M} + \delta \Lambda_1^{M} 
- \Lambda_0^{C} A_2^B - \Lambda_1^B A_1^{C} - \Lambda_0^{C}\delta\Lambda_1^B  \\ \nonumber
A_2^{C} & \rightarrow & A_2^{C} + \delta \Lambda_1^{C} 
 + \Lambda_0^{M} A_2^B + \Lambda_1^B A_1^{M}  - \Lambda_1^B \delta\Lambda_0^{M}.
\ee
The topological operators constructed from $A_2^B, A_1^M$ and $A_1^C$
\be
U_{B}^{(2)}(M_2) &=& e^{i\pi\oint_{M_2} A_2^B} \\
U_{M}^{(1)}(M_1) &=& e^{i\pi\oint_{M_1} A_1^{M}} \\
U_{C}^{(1)}(M_1) &=& e^{i\pi\oint_{M_1} A_1^{C}}
\ee
are therefore gauge invariant, but the analogous operators constructed from 
$A_1^B, A_2^M$ and $A_2^C$ are neither gauge invariant nor topological.
There are two ways to cure this problem.
The first option is to attach a one-higher-dimensional operator with a boundary on the original operator,
such that its gauge variation cancels that of the lower-dimensional operator, similar to anomaly inflow. 
The operators with the required attachments are given by
\be\label{nongenuineops}
&\tilde{U}_{B}^{(1)}(M_1,M_2) = e^{i\pi\oint_{M_1} A_1^B} e^{i\pi \int_{M_2} A_1^{{M}}\cup A_1^{{C}} }, \nonumber\\
&\tilde{U}_{M}^{(2)}(M_2,M_3) = e^{i\pi\oint_{M_2} A_2^{M}} e^{i\pi \int_{M_3} A_2^B\cup A_1^{{C}} }, \\
&\tilde{U}_{C}^{(2)}(M_2,M_3) =  e^{i\pi\oint_{M_2} A_2^{C}} e^{-i\pi \int_{M_3} A_2^B\cup A_1^{{M}} }\nonumber
\ee
where in each case $\partial M_{d+1}=M_d$. 
These operators are {\em non-genuine} in the sense that they depend on a higher dimensional manifold.
The second option is to dress the operator with an equal dimension topological field theory,
whose gauge variation cancels that of the operator. 
In our case the dressed operators are given by 
\be
\hat{U}_{B}^{(1)}(M_1) &=& {1\over 4} \sum_{\phi_0,\phi_0'\in C^0(M_1,\mathbb{Z}_2)} e^{i\pi\oint_{M_1} 
(A_1^B + \phi_0 A_1^{M} + \phi_0' A_1^{C} + \phi_0 \delta \phi'_0)} \nonumber \\
\hat{U}_{M}^{(2)}(M_2) &=& {1\over 4} \sum_{\phi_0\in C^0(M_2,\mathbb{Z}_2),\phi_1\in C^1(M_2,\mathbb{Z}_2)}  e^{i\pi\oint_{M_2} (A_2^{M} + \phi_0 A_2^B 
+ \phi_1 A_1^{C} + \phi_0\delta\phi_1)  } \\\label{symTFToperators}
\hat{U}_{C}^{(2)}(M_2) &=& {1\over 4} \sum_{\phi_0\in C^0(M_2,\mathbb{Z}_2),\phi_1\in C^1(M_2,\mathbb{Z}_2)} 
e^{i\pi\oint_{M_2} (A_2^{C} + \phi_0 A_2^B + \phi_1 A_1^{M} +\phi_0\delta\phi_1)} \nonumber
\ee
where $\phi_0,\phi_0'$, and $\phi_1$ are cochains transforming appropriately under the gauge transformations of $A_2^B, A_1^{M}$ and $A_1^{C}$.
These operators are {\em non-invertible}.

In fact the two versions of the gauge invariant operators are related.
Take for example the operator $\hat{U}_B^{(1)}(M_1)$. Performing the sum over $\phi_0'$ imposes 
\be
\label{phi0condition}
\delta\phi_0 = A_1^C \,,
\ee
which implies that $A_1^C$ has a holonomy which vanishes mod 2 on any 1-cycle $M_1$,
\be
\oint_{M_1} A_1^{C}=0\,\,\text{mod}\,\, 2 \,.
\ee
This requirement is implemented by the insertion of the projector ${1\over 2} (1+ e^{i\pi \oint_{M_1} A_1^{C}})$.
Furthermore the condition in eq. (\ref{phi0condition}) allows us to reexpress the remaining sum over $\phi_0$ as
\be
\sum_{\phi_0\in C^0(M_1, \mathbb{Z}_2)} e^{i\oint_{M_1} \phi_0 A_1^{M}} = \sum_{\chi_0\in H^0(M_1,\mathbb{Z}_2)} 
e^{i\pi \oint_{M_1} \chi_0 A_1^M} e^{i\pi\int_{M_2} A_1^M A_1^C}
\ee
where $\partial M_2 = M_1$. We therefore find that
\be
\label{tildeUBhatUB}
\hat{U}_{B}^{(1)}(M_1) &=& \frac{1}{8}e^{i\pi \oint_{M_1} A_1^B} e^{i\pi \int_{M_2} A_1^{M} A_1^{{C}}}(1+ e^{i\pi \oint_{M_1} A_1^{{M}}}) 
(1+ e^{i\pi \oint_{M_1}A_1^{{C}}}) \nonumber \\
&=& \frac{1}{8} \tilde{U}_{B}^{(1)}(M_1,M_2)(1+U_{M}^{(1)}(M_1))(1+U_{C}^{(1)}(M_1)) \,.
\ee
There are similar relations for the other operators:\footnote{The normalization was discussed in \cite{Kaidi:2021xfk, Roumpedakis:2022aik}.}
\be
\label{tildeUMhatUM}
\hat{U}_{M}^{(2)}(M_2) &=& \frac{1}{4\sqrt{2}}\tilde{U}_{M}^{(2)}(M_2,M_3)\frac{1+U_{B}^{(2)}(M_2)}{|H^0(M_2, \mathbb{Z}_2)|}\sum_{\gamma_1\in H_1(M_2, \mathbb{Z}_2)}U_{C}^{(1)}(\gamma_1) \\
\label{tildeUChatUC}
\hat{U}_{C}^{(2)}(M_2) &=& \frac{1}{4\sqrt{2}}\tilde{U}_{C}^{(2)}(M_2,M_3)\frac{1+U_{B}^{(2)}(M_2)}{|H^0(M_2, \mathbb{Z}_2)|}\sum_{\gamma_1\in H_1(M_2, \mathbb{Z}_2)}U_{M}^{(1)}(\gamma_1).
\ee

\subsection{Link pairings}

\begin{figure}[h!]
\centering
\includegraphics[scale=0.16, trim={4cm, 1.5cm, 2cm, 0.3cm}, clip]{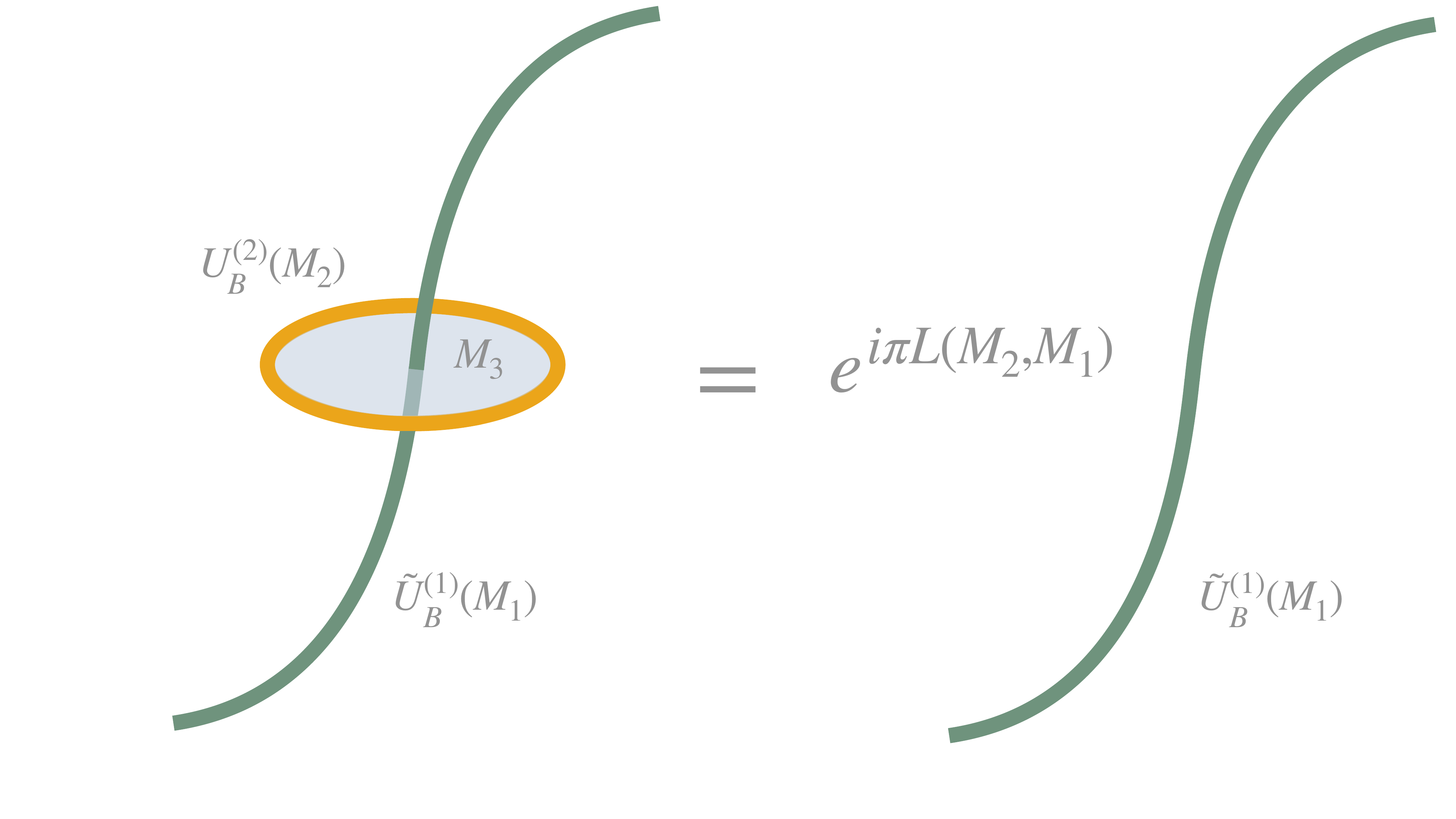}
\caption{Linking between $\tilde{U}_B^{(1)}(M_1)$ and $U_B^{(2)}(M_2)$.}
\label{LinkingStd}
\end{figure}

The action of a topological operator on another topological operator is determined by their link pairing.
In four dimensions a surface operator can link with a line operator, Fig.~\ref{LinkingStd}.
The non-trivial pairings follow directly from the $BF$ terms in the symmetry theory, and are given by
\begin{align}
\label{BBlinking}
&\langle {U}_B^{(2)}(M_2) \tilde{U}_B^{(1)}(M_1,M_2')\rangle = (-1)^{L(M_2,M_1)} \\
&\langle \tilde{U}_M^{(2)}(M_2,M_3) U_M^{(1)}(M_1)\rangle = (-1)^{L(M_2,M_1)} \\
&\langle\tilde{U}_C^{(2)}(M_2,M_3) U_C^{(1)}(M_1)\rangle =  (-1)^{L(M_2,M_1)} \\
\label{MBlinking}
&\langle \tilde{U}_{M}^{(2)}(M_2,M_3) \tilde{U}_{B}^{(1)}(M_1,M_2')\rangle = (-1)^{\frac{\alpha}{2} L(M_2,M_1)} \\
\label{CBlinking}
&\langle \tilde{U}_{C}^{(2)}(M_2,M_3) \tilde{U}_{B}^{(1)}(M_1,M_2')\rangle  =  (-1)^{\frac{\beta}{2} L(M_2,M_1)}
\end{align}
where $L(M_2, M_1)$ is the linking number between the surface $M_2$ and the line $M_1$.
These remain the same if we replace the non-genuine operators $\tilde{U}$ with their non-invertible versions $\hat{U}$.
The link pairings can be equivalently described in terms of the canonical commutators (suppressing delta functions):
\be
\label{Commutators1}
&[A_2^B, A_1^B]=[A_2^{{M}}, A_1^{{M}}]=[A_2^{{C}}, A_1^{{C}}]=\frac{i}{\pi}\\
\label{Commutators2}
&[A_2^{{M}}, A_1^B]= \frac{i\alpha}{2\pi},\,\,\,\, [A_2^{C}, A_1^B]=\frac{i\beta}{2\pi}. 
\ee
The cubic term in the symmetry theory implies that there is a also a triple-link involving the $B$-line operator 
and the $M$ and $C$-surface operators:\footnote{This can be shown for the non-invertible operators $\hat{U}$ following 
\cite{Kaidi:2023maf}. The non-genuine operators $\tilde{U}$ do not appear to have a trivial triple-link.}
\be
\label{triplelinking}
\langle \hat{U}_{B}^{(1)}(M_1)\hat{U}_{{M}}^{(2)}(M_2) \hat{U}_{{C}}^{(2)}(M_2')\rangle=(-1)^{L(M_1, M_2,M_2')}.
\ee

Let us derive two of the link-parings explicitly.
For example, to compute the correlator in eq. (\ref{BBlinking}) we insert a source into the equation of motion for $A_1^B$, which becomes
\be
\label{B1Equation}
\delta A_2^B = \delta^{(3)}(\perp \, M_1) \,.
\ee
This gives
\be 
\oint_{M_2} A_2^B  = \int_{M_3} \delta A_2^B = \int_{M_3} \delta^{(3)}(\perp \, M_1) = I(M_3,M_1) 
=  L(M_2,M_1)  \,,
\ee
where $\partial M_3 = M_2$, and $I(M_3,M_1)$ is the intersection number of $M_3$ and $M_1$, which is the
same as the linking number of $M_2$ and $M_1$.
It follows that the correllator is given by 
\be
\langle {U}_B^{(2)}(M_2) \tilde{U}_B^{(1)}(M_1,M_2')\rangle =
\langle  e^{i\pi\oint_{M_2} A_2^B}  e^{i\pi\oint_{M_1} A_1^B} e^{i\pi \int_{M_2'} A_1^{{M}}\cup A_1^{{C}} }  \rangle 
= (-1)^{L(M_2,M_1)} \,.
\ee
As a more intricate example, take the correlator in eq. (\ref{MBlinking}).
We again insert a source for $A_1^B$, but we look at the equation of motion for $A_1^M$,
\be
\label{A1MEquation}
\delta A_2^M &=& \mbox{} - {\alpha\over 2} \delta A_2^B - A_2^B A_1^C \\
&=& \mbox{} - {\alpha\over 2} \delta^{(3)}(\perp \, M_1) - A_2^B A_1^C \,,
\ee
where in the second equality we used (\ref{B1Equation}). It follows that
\be
 \oint_{M_2} A_2^M  +  \int_{M_3} A_2^B A_1^C &=& \int_{M_3}( \delta A_2^M  + A_2^B A_1^C) 
=  \mbox{} - {\alpha\over 2} L(M_2,M_1) \,,
\ee
which then gives the correlator in eq. (\ref{MBlinking}).
The other link pairings are obtained in a similar way.

\subsection{Fusion}
\label{SectionFusion}

Using a similar approach we can compute the fusion rules for the topological operators.
This will have implications for the algebra of the symmetry operators and for the decomposition of products of the irreducible
representations of the corresponding symmetry.

\begin{figure}[h!]
\centering
\includegraphics[scale=0.235, trim={2cm, 0.1cm, 2cm, 0.1cm}, clip]{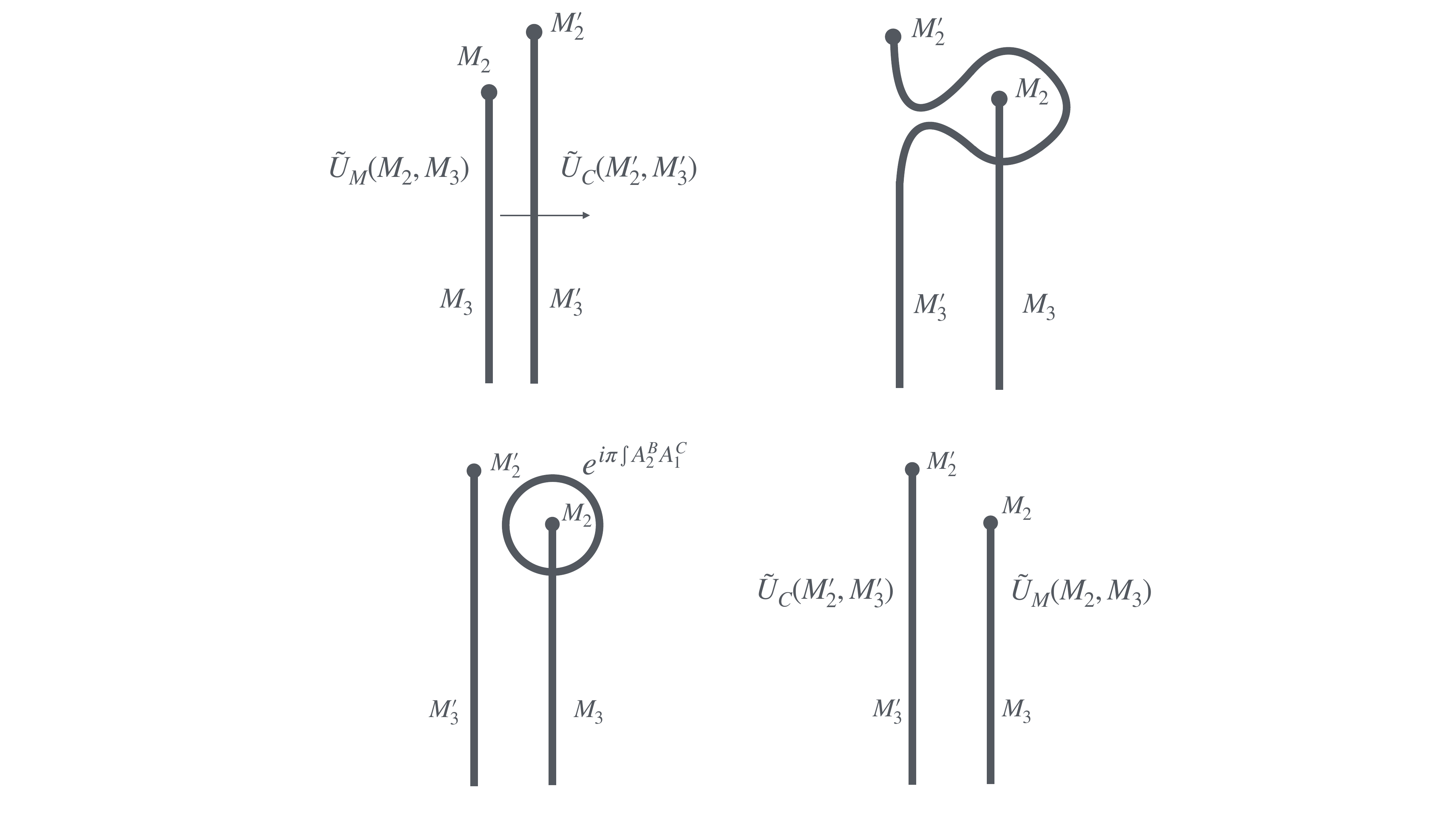}
\caption{Two non-genuine surface operators $\tilde{U}_{M}^{(2)}$ and $\tilde{U}_{{C}}^{(2)}$ do not commute due to the three-surface attached to them.}
\label{LinkingvsIntersection}
\end{figure}

Let us begin with the surface operators.
The self-fusions are given by 
\begin{align}
\label{SurfaceSelfFusion}
&U_B^{(2)}(M_2) U_B^{(2)}(M_2) = 1 \nonumber \\
&\tilde{U}_M^{(2)}(M_2,M_3) \tilde{U}_M^{(2)}(M_2,M_3) = [U_B^{(2)}(M_2)]^\alpha  \\
&\tilde{U}_C^{(2)}(M_2,M_3) \tilde{U}_C^{(2)}(M_2,M_3)  = [U_B^{(2)}(M_2)]^\beta \,. \nonumber
\end{align}
The second relation follows from the equation of motion for $A_1^M$ (\ref{A1MEquation}), and the third
from the analogous equation of motion for $A_1^C$.
In addition, the fusions of different surface operators satisfy
\begin{align}
\label{SurfaceMixedFusion}
&\tilde{U}_M^{(2)}(M_2,M_3) {U}_B^{(2)}(M_2) 
 = {U}_B^{(2)}(M_2) \tilde{U}_M^{(2)}(M_2,M_3) \nonumber \\
&\tilde{U}_C^{(2)}(M_2,M_3) {U}_B^{(2)}(M_2) 
 = {U}_B^{(2)}(M_2) \tilde{U}_C^{(2)}(M_2,M_3) \\
& \tilde{U}_M^{(2)}(M_2,M_3) \tilde{U}_C^{(2)}(M_2,M_3) 
 = U_B^{(2)}(M_2) \tilde{U}_C^{(2)}(M_2,M_3) \tilde{U}_M^{(2)}(M_2,M_3) \nonumber \,.
\end{align}
Note that the $M$-surface operator and the $C$-surface operator fail to commute by a $B$-surface operator.
Naively one would expect surface operators to commute in four dimensions.
However these surface operators are not genuine, since they depend on the choice of a 3d manifold $M_3$.
The presence of the 3-surface operator on $M_3$ is what leads to the non-trivial commutation relation.\footnote{A similar effect leads to non-trivial braiding of line operators in $\mathfrak{su}(N)_k$ YM-CS theories \cite{Argurio:2024oym}.}
The product of the $M$-surface operator and the $C$-surface operator is given by
\be
\tilde{U}_M^{(2)}(M_2,M_3) \tilde{U}_C^{(2)}(M_2,M_3') &=& e^{i\pi\left(\oint_{M_2} A_2^M + \int_{M_3} A_2^B A_1^C\right)}
e^{i\pi\left(\oint_{M_2} A_2^C + \int_{M_3'} A_2^B A_1^M\right)} \nonumber \\
&=& e^{i\pi \int_{M_3}\left(\delta A_2^M + A_2^B A_1^C\right)}
e^{i\pi\left(\oint_{M_2} A_2^C + \int_{M_3'} A_2^B A_1^M\right)} \,.
\ee
Note that we have taken the attached 3-surface to be different for the two operators in order to regularize the computation.
The second operator contains a source term for the equation of motion for $A_1^M$,
\be
\delta A_2^M + A_2^B A_1^C = A_2^B \, \delta^{(1)}(\perp  M_3')  \,.
\ee
Integrating over $M_3$ gives (see fig. \ref{LinkingvsIntersection})
\be
\int_{M_3} \delta A_2^M = \int_{M_3} A_2^B \, \delta^{(1)}(\perp  M_3') = \int_{M_3\cap M_3'} A_2^B = \int_{M_2} A_2^B \,,
\ee
and therefore taking the first operator through the second operator we get
\be
\tilde{U}_M^{(2)}(M_2,M_3) \tilde{U}_C^{(2)}(M_2,M_3')  
 &=& e^{i\pi \int_{M_2} A_2^B} \tilde{U}_C^{(2)}(M_2,M_3') \tilde{U}_M^{(2)}(M_2,M_3) \nonumber \\
&=& U_B^{(2)}(M_2) \tilde{U}_C^{(2)}(M_2,M_3') \tilde{U}_M^{(2)}(M_2,M_3) \,.
\ee
It also follows from this that the self-fusion of the operator 
$\tilde{U}_{MC}^{(2)} \equiv \tilde{U}_M^{(2)}\tilde{U}_C^{(2)}$ is given by
\be
\label{UMCsquared}
\tilde{U}_{MC}^{(2)}(M_2,M_3) \tilde{U}_{MC}^{(2)}(M_2,M_3) = [U_B^{(2)}(M_2)]^{1+ \alpha + \beta} \,.
\ee

The self-fusions of the line operators are given by
\begin{align}
\label{LineSelfFusion}
&U_M^{(1)}(M_1) U_M^{(1)}(M_1) = 1 \nonumber \\
&U_C^{(1)}(M_1) U_C^{(1)}(M_1) = 1 \\
&\tilde{U}_B^{(1)}(M_1,M_2) \tilde{U}_B^{(1)}(M_1,M_2) = [U_M^{(1)}(M_1)]^\alpha [U_C^{(1)}(M_1)]^\beta \nonumber \,,
\end{align}
where the third relation follows from the fact that twice the equation of motion for $A_2^B$ vanishes mod 2,
\be
2\left(\delta A_1^B + A_1^{M} A_1^{C}+ {\alpha\over 2} \delta A_1^{M}+ {\beta\over 2} \delta A_1^{C}\right) =
0 \; \mbox{mod} \; 2 \,.
\ee
The fusions of different line operators are, of course, all commutative.

\subsection{Non-abelian global symmetry}

Let us now consider a specific {\em absolute} theory corresponding to Dirichlet boundary conditions for the three one-form fields
$A_1^B, A_1^M, A_1^C$, and Neumann boundary conditions for the three 2-form fields
$A_2^B, A_2^M, A_2^C$.
 This theory has a 0-form global symmetry that is generated by the three surface operators $U_B^{(2)}, \tilde{U}_M^{(2)}, \tilde{U}_C^{(2)}$,
taken to the topological boundary. The line operators $\tilde{U}_B^{(1)}, U_M^{(1)}, U_C^{(1)}$ are trivial at the boundary due to the boundary conditions.
On the other hand if we orient these line operators along the extra dimension of the SymTFT, their endpoints on the boundary 
correspond to nontrivial local operators in the boundary theory.
The link-pairings found above imply that these operators are charged under the global symmetry generated by the surface operators.
In other words, the line operators should correspond to representations of the global symmetry generated by the surface operators.

The fusion rules of the surface operators in eqs. (\ref{SurfaceSelfFusion}), (\ref{SurfaceMixedFusion}) are precisely the algebra of either $\mathbb{D}_8$ or $\mathbb{Q}_8$ (see Appendix \ref{Groups} for 
the definition and properties of these groups). In standard notation, the generators of either group are denoted by $a$ and $x$.
The two groups are defined as 
\be
\mathbb{D}_8 &=& \{ a,x: a^4=1, \, x^2=1, \, xax^{-1}= a^3 \} \\
\mathbb{Q}_8 &=& \{a,x: a^4=x^4=1, \, a^2=x^2, \, xax^{-1}=a^3\} \,.
\ee
The order 2 element $a^2$ is identified with the $B$-surface operator $U_B^{(2)}$. The 
elements
$a, x, xa$ are identified with $\tilde{U}_M^{(2)}, \tilde{U}_C^{(2)}, \tilde{U}_{MC}^{(2)}$ in a way that depends on the values of $\alpha$ and $\beta$.
If $\alpha= \beta = 1$ we see from eqs. (\ref{SurfaceSelfFusion}) 
and (\ref{UMCsquared}) that
all three surface operators are order 4 and the group is $\mathbb{Q}_8$.
The other three cases give $\mathbb{D}_8$, where the single order 4 element $a$ is identified as follows
\be 
a = \left\{
\begin{array}{ll}
\tilde{U}_M^{(2)} & (\alpha,\beta) = (1,0) \\
\tilde{U}_C^{(2)} & (\alpha,\beta) = (0,1) \\
\tilde{U}_{MC}^{(2)} & (\alpha,\beta)=(0,0) \,.
\end{array}
\right.
\ee

We now turn our attention to the three line operators $\tilde{U}_B^{(1)}, U_M^{(1)}, U_C^{(1)}$.
These should correspond to irreducible representations of the global symmetry.
Their fusion rules should therefore describe the decomposition of the products of these representations.
Other than the trivial representation, both $\mathbb{D}_8$ and $\mathbb{Q}_8$ have three 1d representations,
and one 2d representation (see Appendix \ref{Groups}).
The trivial self-fusion of $U_M^{(1)}$ and $U_C^{(1)}$ in eq. (\ref{LineSelfFusion}), as well as the trivial self-fusion of $U_{MC}^{(1)} = U_M^{(1)} U_C^{(1)}$ 
that follows from these, identifies these three with the three nontrivial 1d representations of $\mathbb{D}_8$ or $\mathbb{Q}_8$.
The remaining line operator $\tilde{U}_B^{(1)}$ should correspond to the 2d representation.
However its self-fusion in eq. (\ref{LineSelfFusion}) does not have the form of the decomposition of the product of two 2d representations in eq. (\ref{2dtimes2d}).
On the other hand the non-invertible version of the line operator $\hat{U}_B^{(1)}$ in eq. (\ref{tildeUBhatUB}) does give the correct decomposition,
\be
\label{RepDecomposition}
\hat{U}_B^{(1)}(M_1) \hat{U}_B^{(1)}(M_1) &=& {1\over 4} \left(1 + U_M^{(1)}(M_1) + U_C^{(1)}(M_1) + U_{MC}^{(1)}(M_1)  \right) ,
\ee
namely the direct sum of the 1d representations.
Furthernore, its fusion with the line operators corresponding to the 1d representations is given by
\begin{align}
&\hat{U}_B^{(1)}(M_1) U_M^{(1)}(M_1) = \hat{U}_B^{(1)}(M_1) \nonumber \\
&\hat{U}_B^{(1)}(M_1) U_C^{(1)}(M_1)= \hat{U}_B^{(1)}(M_1) \\
&\hat{U}_B^{(1)}(M_1) U_{MC}^{(1)}(M_1) = \hat{U}_B^{(1)}(M_1) , \nonumber
\end{align}
which is also consistent with its identification with the 2d representation.

\section{Explicit realization in 3d gauge theory}

\label{orthosymplecticanalysis}

\subsection{Orthogonal gauge theory} 
A simple class of theories possessing this structure is given by an $\mathfrak{so}(2n)$ gauge theory with an even-level 
Chern-Simons coupling $\kappa = 2k$ \cite{Cordova:2017vab}.

The theory in which the global form of the gauge group is $SO(2n)$ has a global symmetry 
$\mathbb{Z}_2^{(1)} \times \mathbb{Z}_{2,{M}}^{(0)} \times \mathbb{Z}_{2,C}^{(0)}$,
where
$\mathbb{Z}_2^{(1)}$ is the electric one-form symmetry,
$\mathbb{Z}_{2,{M}}^{(0)}$ is the magnetic 0-form symmetry,
and $\mathbb{Z}_{2,{C}}^{(0)}$ is the outer-automorphism of the $\mathfrak{so}(2n)$ algebra (except for $n= 4$, where it is $S_3$), 
namely charge conjugation.
The operators charged under the 1-form symmetry are Wilson lines.
A Wilson line in the vector representation of $SO(2n)$ carries one unit of charge under $\mathbb{Z}_2^{(1)}$.
An even number of such Wilson lines can be screened by gluons.
The operators charged under the magnetic 0-form symmetry are monopoles.
The monopole operator in the $SO(2n)$ gauge theory corresponds to an $SO(2n)$ bundle with  a non-trivial second Stieffel-Whitney class $w_2\neq 0$.
If $k\neq 0$ the monopole operator is not gauge invariant. 
It transforms in the $2k$-fold symmetric product of the vector representation of $SO(2n)$.\footnote{Note that for an odd CS level, 
the monopole would screen the Wilson line, and therefore break the $\mathbb{Z}_2$ 1-form symmetry.}
There are no gauge invariant local operators charged under charge conjugation. 

The symmetries of $SO(2n)_{2k}$ theory have the mixed anomalies described in section \ref{symTFTD8Q8}, 
as well as a self-anomaly for the 1-form symmetry \cite{Cordova:2017vab}. The anomaly theory is given by
\be
\label{AnomalyAction1}
S_4^{\text{anomaly}} 
=  i\pi \int_{M_4} A_2^B \cup \left( {n\over 2} \delta A_{1}^{M} + {k\over 2} \delta A_{1}^{C} + A_{1}^{M} \cup A_{1}^{C}  + nk A_2^B 
\right) .
\ee
The 1-form symmetry anomaly is nontrivial when both $n$ and $k$ are odd.
This is related to the fact that the operator generating the 1-form symmetry has spin $\frac{nk}{4}$, and the symmetry can be gauged if and only if the spin is half-integer (on Spin manifolds). 
If either $n$ or $k$ or both are even, the 1-form symmetry anomaly is trivial and the anomaly theory is identical to eq. \eqref{GeneralAnomalyAction},
with $\alpha = n, \beta = k$, and $\gamma = 1$.
In these three cases one can gauge the 1-form symmetry.
This changes the global structure of the gauge group to $PSO(2n)_{2k}$, and as we saw previously gives a $\mathbb{D}_8$ global 0-form symmetry.
But there is no version of the orthogonal gauge theory with a $\mathbb{Q}_8$ global symmetry, since we cannot gauge the 1-form symmetry 
when both $n$ and $k$ are odd.

\subsection{Orthosymplectic quiver gauge theory} 

To incorporate the $\mathbb{Q}_8$ global symmetry, we consider a class of orthosymplectic quiver gauge theories, namely a 3d gauge theory with 
gauge algebra $\mathfrak{so}(2n)_{2k} \times \mathfrak{usp}(2n)_{-k}$, and matter in the bi-fundamental representation.
We will actually be interested in the maximally supersymmetric version of this theory, which is the ${\cal N}=5$ supersymmetric 
Chern-Simons-Matter theory, since it has a simple holographic dual \cite{Aharony:2008gk}.
But the symmetry structure of the field theory is the same with or without supersymmetry.

In the theory with gauge group $SO(2n)\times USp(2n)$ the global symmetry is the same as for the pure $SO(2n)$ gauge theory, namely
$\mathbb{Z}_2^{(1)} \times \mathbb{Z}_{2,{M}}^{(0)} \times \mathbb{Z}_{2,C}^{(0)}$.
The magnetic and charge conjugation symmetries act only on the $SO(2n)$ sector.
But in this case they act faithfully on gauge invariant local operators.
In particular the monopole operators can be dressed with matter fields into gauge invariant operators.
Denoting the bare $SO(2n)$ monopole operator with unit magnetic flux as $T_1$ and the matter field as $\phi_i^a$, 
the gauge invariant dressed monopole is given by 
\be
\label{DressedMonopole}
{\cal M} = T_1^{(i_1 \cdots i_{k}j_1\cdots j_k)} (\phi^2)_{i_1 j_1} \cdots (\phi^2)_{i_k j_k} \,,
\ee
where $\phi^2_{ij} \equiv J_{ab} \phi_i^a \phi_j^b$, and $J_{ab}$ is the invariant tensor of $USp(2n)$.
This operator carries one unit of charge under $\mathbb{Z}_{2,M}^{(0)}$ and is neutral under $\mathbb{Z}_{2,C}^{(0)}$.\footnote{In the 
${\cal N}=5$ supersymmetric theory this is a non-BPS operator. For example, for $k=1$ it has a classical dimension $\Delta=1$,
and transforms in the ${\bf 5} + {\bf 1}$ of the $R$-symmetry $Spin(5)_R = USp(4)_R$.
The only short superconformal multiplet containing such an operator as its lowest component is the stress tensor multiplet.
But that component is already identified with the meson operator $\mbox{Tr}(\phi^2)$.
The operator in eq. (\ref{DressedMonopole}) must therefore belong to a long multiplet.
A BPS monopole operator is obtained by turning on a magnetic flux in both $SO(2n)$ and $USp(2n)$,
\be 
{\cal M}_{BPS}
= (T_{1,1})^{(i_1 \cdots i_{2k})}_{(a_1 \cdots a_{2k})} \phi^{a_1}_{i_1} \cdots \phi^{a_{2k}}_{i_{2k}} \,.
\ee
This operator carries the same $\mathbb{Z}_{2,M}^{(0)}$ charge as the non-BPS operator in eq. (\ref{DressedMonopole}).}
There is also a baryon-like operator charged under $\mathbb{Z}_{2,C}^{(0)}$ and given by
\be 
{\cal B}
&=& (\phi^2)_{i_1 j_1} \cdots (\phi^2)_{i_N j_N} 
\epsilon^{i_1\cdots i_N j_1\cdots j_N}.
\ee 
The $\mathbb{Z}_2^{(1)}$ 1-form symmetry corresponds to the diagonal combination of the two $\mathbb{Z}_2^{(1)}$ 1-form symmetries, 
associated to the two gauge group factors, which is left unbroken by the bi-fundamental fields.
Equivalently stated, a Wilson line in the $\bold{(2n,1)}$ representation 
is equivalent to a Wilson line in the $\bold{(1,2n)}$ representation via a bi-fundamental field.

The 0-form and 1-form symmetries of the orthosymplectic theory have the same mixed anomalies as the $SO(2n)_{2k}$ theory,
but the anomaly of the 1-form symmetry is absent, since the $USp(2n)_{-k}$ theory has the same anomaly as the $SO(2n)_{2k}$ theory
\cite{Benini:2017dus}.\footnote{The authors of \cite{Benini:2017dus} considered a $USp(2n)_k$ theory with $N_f$ fermions in the vector representation, 
which exhibits an anomaly of the form $w_2 \cup w_2$ when $nk$ is odd and $N_f$ is even. In the present case this becomes
an anomaly for the 1-form symmetry.}  
The anomaly action in this case is therefore 
\be
\label{AnomalyAction1}
S_4^{\text{anomaly}} 
=  i\pi \int_{M_4} A_2^B \cup \left( {n\over 2} \delta A_{1}^{M} + {k\over 2} \delta A_{1}^{C} + A_{1}^{M} \cup A_{1}^{C} \right) \,,
\ee
and the full SymTFT action is
\be
\label{SymTFTAction}
S^{\text{sym}}_4  =  i \pi \int_{M_4} \left[A_2^B  \delta \left(A_1^B + {n\over 2}A_1^M + {k\over 2}A_1^C \right)
+ A_{2}^{M}  \delta A_{1}^{M} + A_{2}^{C} \delta A_{1}^{C} +  A_2^B A_1^M A_1^C \right] .
\ee
Now we can gauge the 1-form symmetry regardless of the values of $n$ and $k$.
The resulting theory has a gauge group $(SO(2n)_{2k}\times USp(2n)_{-k})/\mathbb{Z}_2$. 
The global symmetry is $\mathbb{D}_8$ if either $n$ or $k$ is even, and $\mathbb{Q}_8$ if both $n$ and $k$ are odd.

\subsection{Symmetry webs}

By gauging other subgroups of the global symmetry we get different global forms of the gauge symmetry.\footnote{The subgroups are considered equivalent if related by an inner automorphism of the global symmetry group we start from.} 
A good place to start is with the theory we just discussed, where the gauge group is $(SO(2n)_{2k}\times USp(2n)_{-k})/\mathbb{Z}_2$.
This theory has a purely 0-form global symmetry, which is $\mathbb{D}_8$ if either $n$ or $k$ is even, and $\mathbb{Q}_8$ if both $n$ and $k$ are odd. 
All the other theories can be obtained by gauging subgroups of the 0-form symmetry, producing a {\em symmetry web} of theories 
\cite{Bhardwaj:2022maz}.\footnote{The $\mathbb{D}_8$-web was discussed in \cite{Bhardwaj:2022maz, Bartsch:2022ytj}. 
The $\mathbb{Q}_8$-web is a new result.}

\begin{figure}[h!]
\center
\includegraphics[height=0.55\textwidth]{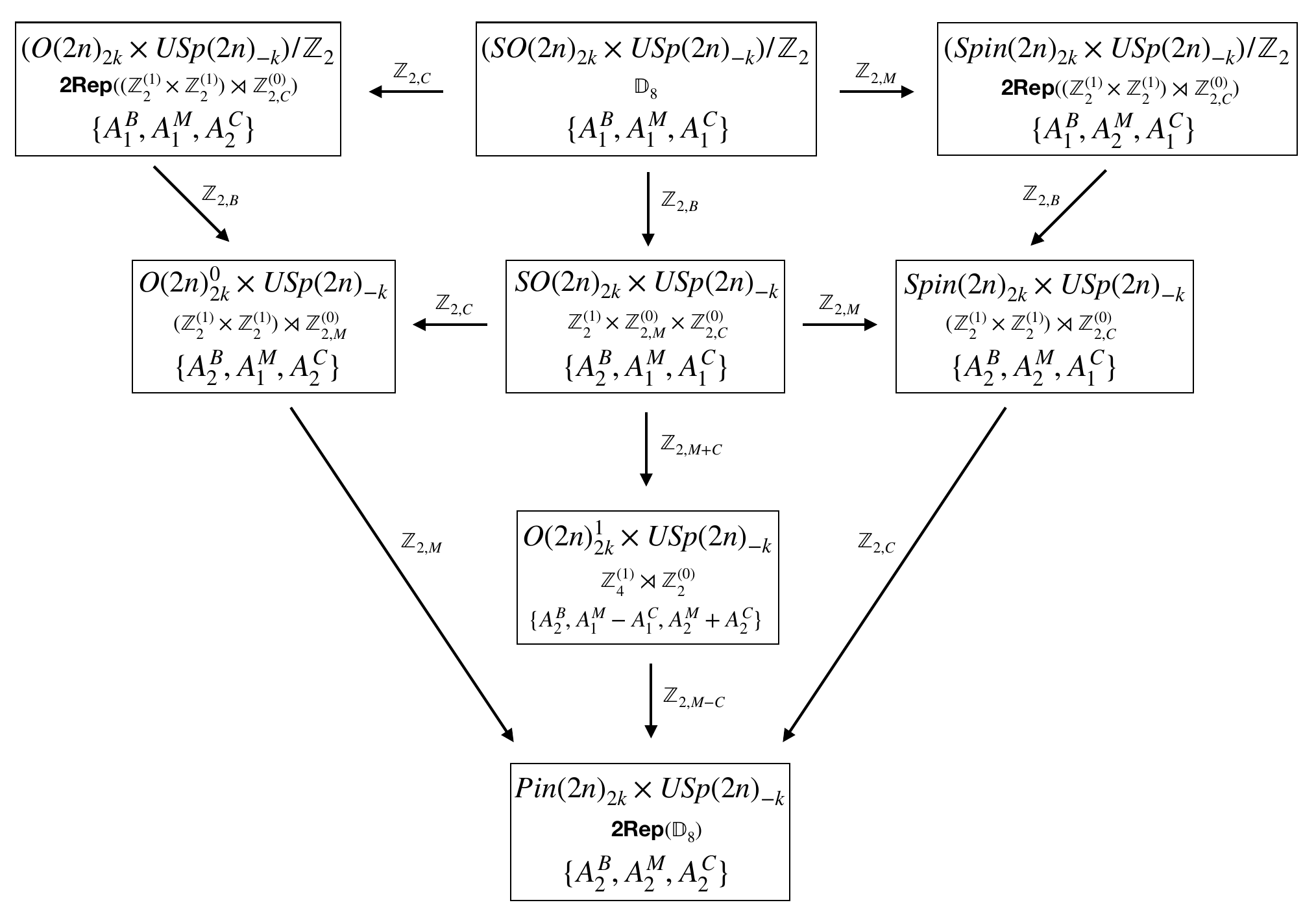}
\caption{The 3d $\mathbb{D}_8$ symmetry web (partly reproduced from \cite{Bhardwaj:2022maz, Bartsch:2022ytj}). Here $n$ and $k$ are both even.
The bulk (SymTFT) fields satisfying Dirichlet BC's are shown in each case.}
\label{WebD8}
\end{figure}

In the case of $\mathbb{D}_8$ there are three $\mathbb{Z}_2$ subgroups 
$\mathbb{Z}_2^x, \mathbb{Z}_2^{ax}, \mathbb{Z}_2^{a^2}$ generated by the order two elements of $\mathbb{D}_8$, 
two $\mathbb{Z}_2^2$ subgroups $\mathbb{Z}_2^{a^2}\times \mathbb{Z}_2^x, \mathbb{Z}_2^{a^2}\times \mathbb{Z}_2^{ax}$,
and a $\mathbb{Z}_4$ subgroup $\mathbb{Z}_4^{a}$ generated by the order four element $a$.
Since the global symmetry is anomaly free, we can gauge any of these subgroups or the full $\mathbb{D}_8$.
This produces a total of eight different theories, as shown in Fig.~\ref{WebD8}.
For example, gauging $\mathbb{Z}_2^{a^2}$ leads to a theory with a global symmetry 
$\mathbb{Z}_2^{(1)}\times \mathbb{Z}_2^{(0)} \times \mathbb{Z}_2^{(0)}$.
This is the theory with gauge group $SO(2n)_{2k}\times USp(2n)_{-k}$, which is what we started with previously. 
This theory has a mixed anomaly involving the $\mathbb{Z}_2^{(1)}$ 1-form symmetry and the two $\mathbb{Z}_2^{(0)}$ symmetries.
Gauging the 1-form symmetry leads back to the $(SO(2n)_{2k}\times USp(2n)_{-k})/\mathbb{Z}_2$ theory, and the global symmetry is extended to $\mathbb{D}_8$.
On the other hand gauging a $\mathbb{Z}_2^{(0)}$ factor leads to a 2-group extension, as seen in the $Spin(2n)_{2k}\times USp(2n)_{-k}$
and $O(2n)_{2k}^{0,1}\times USp(2n)_{-k}$ theories.\footnote{The 2-group is either $(\mathbb{Z}_2^{(1)}\times \mathbb{Z}_2^{(1)})\rtimes \mathbb{Z}_2^{(0)}$
or $\mathbb{Z}_4^{(1)} \rtimes \mathbb{Z}_2^{(0)}$, depending on the theory and on the parity of $n$ and $k$.}
From this point, gauging another $\mathbb{Z}_2$ symmetry leads to a non-invertible global symmetry.
For example, gauging the remaining $\mathbb{Z}_2^{(0)}$ symmetry of the $Spin(2n)_{2k}\times USp(2n)_{-k}$ theory
leads to the $Pin^+(2n)_{2k}\times USp(2n)_{-k}$ theory, which has a non-invertible symmetry given by the {2-}category 
${\bf{2{Rep}}}(\mathbb{D}_8)$.
This is also the result of gauging the full $\mathbb{D}_8$ symmetry of the $(SO(2n)_{2k}\times USp(2n)_{-k})/\mathbb{Z}_2$ theory.

\begin{figure}[h!]
\center
\includegraphics[height=0.55\textwidth]{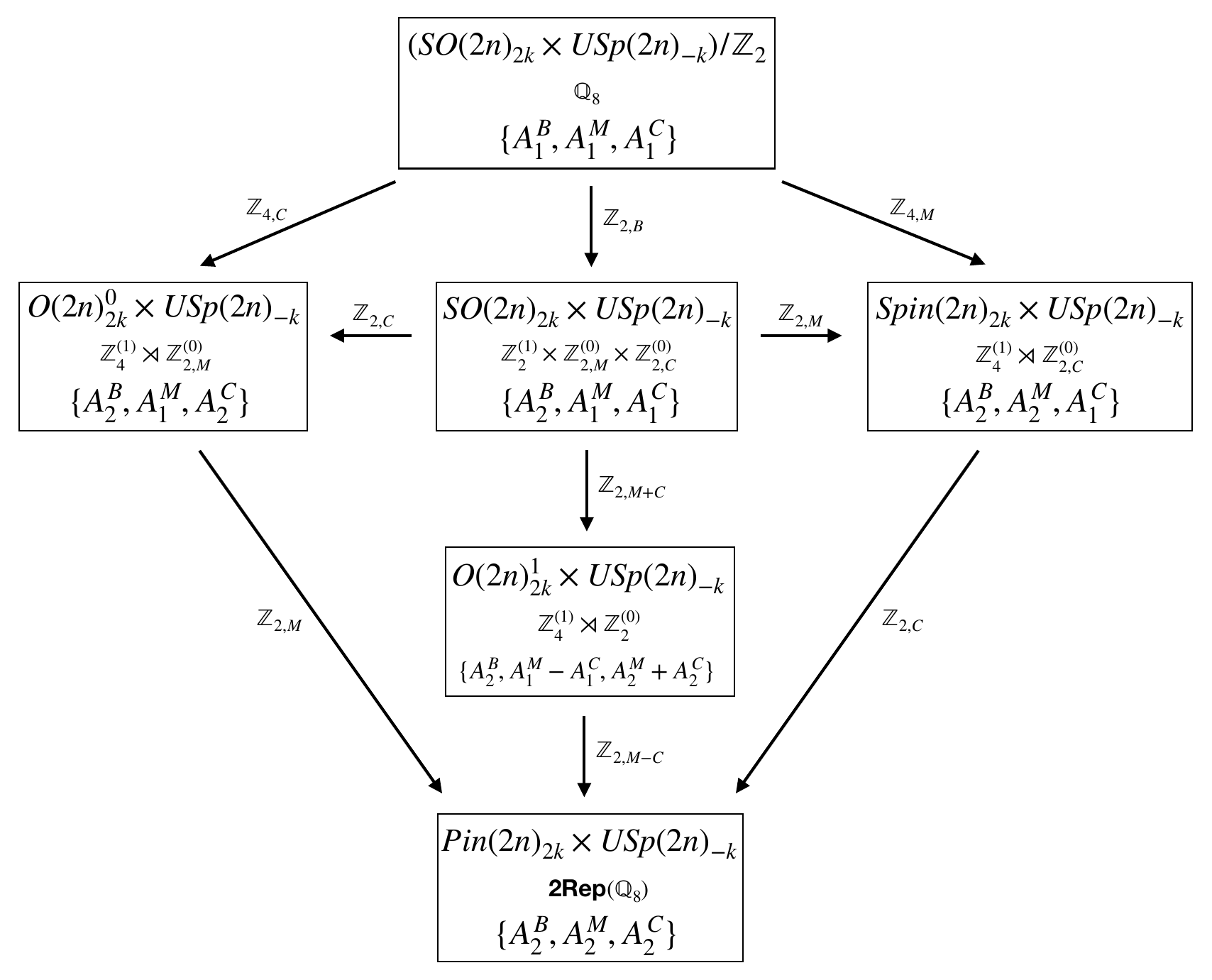}
\caption{The 3d $\mathbb{Q}_8$ symmetry web. Here both $n$ and $k$ are odd.}
\label{WebQ8}
\end{figure}

In the case of $\mathbb{Q}_8$ there are only four subgroups:
$\mathbb{Z}_2^{a^2}$ generated by the element $a^2=x^2=(ax)^2$, and three $\mathbb{Z}_4$ subgroups $\mathbb{Z}_4^a, \mathbb{Z}_4^x, \mathbb{Z}_4^{ax}$.
The symmetry web therefore has only six different theories, Fig.~\ref{WebQ8}.

Alternatively, we can describe the symmetry web in terms of the boundary conditions on the fields of the SymTFT.
The allowed boundary conditions are constrained by the link-parings in eqs. (\ref{BBlinking})-(\ref{CBlinking}), or equivalently by the commutation relations in eqs. (\ref{Commutators1})-(\ref{Commutators2}), that follow from the $BF$ terms in the SymTFT.
The set of fields that can satisfy a Neumann boundary condition must be mutually commuting, and we must choose a maximal set.
The remaining fields must satisfy a Dirichlet boundary condition.
This was referred to in \cite{Witten:1998wy} as a choice of {\em polarization}.
There is an additional condition coming from the triple-link in eq. (\ref{triplelinking}), which forbids the fields involved in the
cubic term of the SymTFT, namely $A_2^B,A_1^M$ and $A_1^C$, from simultaneously satisfying Neumann boundary conditions.
The set of fields satisfying a Dirichlet boundary condition in each case is also shown in Figs.~\ref{WebD8},\ref{WebQ8}.

\section{Holography}
\label{stringtheory}

The ${\cal N}=5$ supersymmetric version of the 3d orthosymplectic quiver theory has a known holographic dual,
given by M theory on $AdS_4 \times S^7/\hat{\mathbb{D}}_k$, or equivalently by Type IIA string theory on $AdS_4\times \mathbb{C}P^3/\mathbb{Z}_2$
\cite{Aharony:2008gk}. 
We will focus on the string theory dual (which strictly speaking is a valid description for $n^{1/5} \ll k \ll n$).
In the Type IIA string theory dual the internal space $\mathbb{C}P^3/\mathbb{Z}_2$ is obtained by quotienting $\mathbb{C}P^3$ by
the following action on the homogeneous coordinates
\be
(z_1, z_2, z_3, z_4)\rightarrow (iz_2^\ast, -iz_1^\ast, iz_4^\ast, -iz_3^\ast) \,,
\ee
combined with the action of $\Omega (-1)^{F_L}$ on the worldsheet.
The latter acts on the Type IIA gauge fields as 
\begin{align}
\label{Z2Action}
B_2  \rightarrow & - B_2 \nonumber \\
B_6  \rightarrow &\,  B_6 \nonumber \\
C_{4n+1}& \rightarrow  - C_{4n+1} \nonumber \\
C_{4n+3} &\rightarrow   C_{4n+3}  \,.
\end{align}
The 4d effective bulk theory in $AdS_4$ will be obtained by reducing Type IIA supergravity on this space,
and the relevant 4d gauge fields will arise from reducing the 10d gauge fields on its homology cycles.
For the fields that are odd under $\Omega (-1)^{F_L}$ we will need to use homology with local coefficients (twisted homology).
The homology groups of $\mathbb{C}P^3/\mathbb{Z}_2$ are shown in Table~\ref{HomologyGroups} (see Appendix~\ref{HomologyAppendix} for more details).
In particular the integer-valued even-dimensional cycles descend from those of $\mathbb{C}P^3$, and correspond to $\mathbb{C}P^k\subset \mathbb{C}P^3$.

The Type IIA background dual to the $\mathfrak{so}(2n)_{2k} \times \mathfrak{usp}(2n)_{-k}$ theories
has RR fluxes on the 6-cycle and 2-cycle given by 
\be
{1\over 2\pi} \int_{\mathbb{C}P^3} F_6 = 2n  \quad , \quad {1\over 2\pi} \int_{\mathbb{C}P^1} F_2 = 2k \,.
\ee
More generally, the Type IIA background may include the additional fluxes 
\be
{1\over 2\pi} \int_{\mathbb{C}P^1} B_2 &=& {\ell \over 2k} \qquad (0\leq \ell \leq k-1) \\
{1\over 2\pi} F_0 &=& \Delta k  \,,
\ee
corresponding respectively to the realtive rank and relative level in the 
$\mathfrak{so}(2(n+\ell))_{2(k+\Delta k)} \times \mathfrak{usp}(2n)_{-k}$ theories,
as well as discrete torsion in the NSNS or RR sector,
\be
\theta_{NS} &=& {1\over 2\pi} \int_{\tilde{\Sigma}_2} B_2 \\
\theta_{RR} &=& {1\over 2\pi} \int_{\Sigma_3} C_3 \,,
\ee
where $\tilde{\Sigma}_2$ is the twisted torsion 2-cycle in $H_2(\mathbb{C}P^3/\mathbb{Z}_2,\tilde{\mathbb{Z}}) = \mathbb{Z}_2$,
and $\Sigma_3$ is the torsion 3-cycle in $H_3(\mathbb{C}P^3/\mathbb{Z}_2,\mathbb{Z}) = \mathbb{Z}_2$.
These correspond, respectively, to changing $\mathfrak{so}(2n)$ to $\mathfrak{so}(2n+1)$, and to exchanging the 
$\mathfrak{so}$ and $\mathfrak{usp}$ factors.\footnote{This also clarifies a point which was a little obscure in \cite{Aharony:2008gk}, where 
no distinction was made between the integer holonomy of $B_2$ on $\mathbb{C}P^1$ and the torsion holonomy 
of $B_2$ on $\tilde{\Sigma}_2$. Accounting separately for the $k$ possible values of the former and the two possible
values of the latter, together with the two possible values of RR torsion, gives $4k$ possibilities, in agreement with the counting
from the M-theory viewpoint in which there is only discrete torsion taking values in 
$H_3(S^7/\hat{\mathbb{D}}_k,\mathbb{Z}) = \mathbb{Z}_{4k}$.}

The different background fluxes are sourced by domain walls corresponding to branes wrapping various cycles
of the internal space \cite{Bergman:2010xd}. The overall rank $n$ corresponds to unwrapped D2-branes, the relative rank $\ell$ to
D4-branes wrapping $\mathbb{C}P^1$, the overall level $k$ to D6-branes wrapping $\mathbb{C}P^2$,
and the relative level to D8-branes wrapping the entire $\mathbb{C}P^3/\mathbb{Z}_2$.
The discrete NSNS torsion $\theta_{NS}$ is sourced by an NS5-brane wrapping the torsion 3-cycle $\Sigma_3$,
and the discrete RR torsion $\theta_{RR}$ is sourced by a D4-brane wrapping the twisted torsion 2-cycle $\tilde{\Sigma}_2$.

Since in this paper we are focussing on the $\mathfrak{so}(2n)_{2k} \times \mathfrak{usp}(2n)_{-k}$ theories,
we will not consider the additional fluxes.

\begin{table}[h]
\centering
{\renewcommand{\arraystretch}{1.2}
\begin{tabular}{ |c|c|c|c|c|c|c|c|} 
\hline
$p$ & 0 & 1 & 2 & 3 & 4 & 5 & 6 \\ 
 \hline
$H_p(\mathbb{C}P^3/\mathbb{Z}_2,\mathbb{Z})$ & $\mathbb{Z}$ & $\mathbb{Z}_2$ & 0 & $\mathbb{Z}_2$ &  $\mathbb{Z}$ & $\mathbb{Z}_2$  & 0 \\ 
$H^p(\mathbb{C}P^3/\mathbb{Z}_2,\mathbb{Z})$ & $\mathbb{Z}$ & 0  & $\mathbb{Z}_2$ & 0 &  $\mathbb{Z}\oplus \mathbb{Z}_2$ & 0  & $\mathbb{Z}_2$ \\ 
\hline 
$H_p(\mathbb{C}P^3/\mathbb{Z}_2,\tilde{\mathbb{Z}})$ & $\mathbb{Z}_2$ & 0 & $\mathbb{Z} \oplus \mathbb{Z}_2$ &  0 & $\mathbb{Z}_2$ & 0  & $\mathbb{Z}$ \\ 
$H^p(\mathbb{C}P^3/\mathbb{Z}_2,\tilde{\mathbb{Z}})$ & 0 & $\mathbb{Z}_2$ & $\mathbb{Z}$ &  $\mathbb{Z}_2$ & 0 & $\mathbb{Z}_2$  & $\mathbb{Z}$\\
 \hline
\end{tabular}
}
\caption{Homology and cohomology groups of $\mathbb{C}P^3/\mathbb{Z}_2$.}\label{homologiesCP3Z2}
\label{HomologyGroups}
\end{table}

\subsection{SymTFT from dimensional reduction}

The SymTFT will arise as the topological sector in the dimensional reduction of Type IIA supergravity to $AdS_4$.
The reduction of supergravity to $AdS_4$ is not in general topological, but the topological sector dominates near the conformal boundary,
which represents the topological boundary of the SymTFT.

The relevant part of the Type IIA supergravity action is given by 
\be
\label{IIASUGRA1}
S_{10}  =  \mbox{} -{1\over 2\pi} \int_{M_{10}} d^{10}x \sqrt{-g} \left[ {1\over 2} e^{-2\Phi}  |H_3|^2
+ {1\over 2}|F_2|^2 + {1\over 2}|{F}_4|^2\right] 
 -  {1\over 8\pi^2} \int_{M_{10}} B_2 dC_3 dC_3 \,, \nonumber \\
\ee
where $H_3=dB_2$, $F_2=dC_1$, and ${F}_4 = dC_3 - \frac{1}{2\pi}H_3\wedge C_1$.\footnote{We normalize all the gauge fields to have 
fluxes that are integer multiples of $2\pi$.}
It will be convenient to work in a democratic formulation of Type IIA supergravity, in which the 10d gauge fields and their Hodge duals are treated as 
independent fields.
The action for the gauge fields is expressed as a topological field theory in eleven dimensions that depends only on the field 
strengths \cite{Belov:2006xj,Apruzzi:2023uma}
\be
\label{IIASUGRA2}
S_{11} & = & {1\over 2\pi} \int_{M_{11}} \left[H_3 dH_7 - F_2 dF_8 + F_4 dF_6 + \frac{1}{2\pi}H_3 F_2 F_6 - {1\over 4\pi}H_3 F_4 F_4\right] ,
\ee
where  $M_{11}$ is an 11d manifold whose boundary is $M_{10}$.\footnote{
Strictly speaking, one should add kinetic terms on the boundary $M_{10}$ to have a well-defined variational principle 
\cite{Maldacena:2001ss, Evnin:2023ypu}. Requiring the surface terms to cancel imposes the Hodge duality relations in $M_{10}$.
For us the boundary terms will not be important since they will reduce to the kinetic terms for the 4d fields, 
which are subdominant to the topological terms near the boundary of $AdS_4$.}
In terms of the higher form potentials we have\footnote{Our conventions for the higher
form fields are $F_{10-p} = (-1)^{\lfloor {p\over 2} \rfloor} *F_p$ and $H_7 = e^{-2\Phi} *H_3$.
Note also that $**=(-1)^{1+p(10-p)}$, and that $|F_p|^2 \sqrt{-g} d^{10} x = (-1)^{1+p(10-p)}F_p\wedge *F_p$.}
\be
F_p &=& dC_{p-1} - {1\over 2\pi} H_3 C_{p-3} \\
H_7 &=& dB_6 - {1\over 4\pi} \left(F_6 C_1 - F_4 C_3 +  F_2 C_5\right) \,.
\ee
This action puts the 10d Bianchi identities and equations of motion on an equal footing.
Half of the fields serve as Lagrange multipliers for the Bianchi identities of the other half.
They are all equations of motion of the 11d action:
\be
\label{FpBianchiID}
dF_p &=& \frac{1}{2\pi}H_3 F_{p-2}\\ 
\label{H3BianchiID}
dH_3 &=& 0 \\
\label{H7BianchiID}
dH_7 &=& -\frac{1}{2\pi}F_2 F_6 + {1\over 4\pi} F_4 F_4. 
\ee

The reduction of the 10d theory on $M_4 \times \mathbb{C}P^3/\mathbb{Z}_2$ then corresponds to the reduction of the 11d theory
on $M_5\times \mathbb{C}P^3/\mathbb{Z}_2$, where $\partial M_5 = M_4$.
The reduction process is slightly involved due to the presence of torsion cycles in the internal space. 
We will adopt the approach of \cite{Camara:2011jg} to torsion cohomology, in which a non-trivial class 
$[\alpha_r] \in \mbox{Tor} H^r(X_d,\mathbb{Z}) = \mathbb{Z}_k$ is represented by a pair of non-harmonic eigen-forms of the Laplacian
$(\alpha_r,\omega_{r-1})$ satisfying the relation\footnote{For a more refined treatment using differential cohomology
see \cite{GarciaEtxebarria:2024fuk}.}
\be 
 d\omega_{r-1} =  (-1)^{r-1} k \alpha_r \,.
\ee
The $r-1$ form $\omega_{r-1}$ is globally well defined, and can be integrated over  
an $(r-1)$-cycle. 
This takes values in $\mbox{Tor} H_{r-1}(X_d,\mathbb{Z})$, which is the same as $\mbox{Tor} H^r(X_d,\mathbb{Z})$
by the universal coefficient theorem
\be
\mbox{Tor} H^r(X_d,\mathbb{Z}) \simeq \mbox{Tor} H^{r-1}(X_d,U(1)) \simeq \mbox{Tor}H_{r-1}(X_d,\mathbb{Z}) \,.
\ee

We expand the field strengths in the basis of the cohomologies of $\mathbb{C}P^3/\mathbb{Z}_2$, with integer or local coefficients,
as appropriate for the given 10d field\footnote{We are suppressing the components of the fields that correspond to the 
parameters $\ell$ and $\Delta k$, as well as those corresponding to $\theta_{NS}$ and $\theta_{RR}$,
all of which are assumed to be trivial in this paper.}
\be
\label{expansion}\nonumber
F_2 &=& F_2^{(2)}\tilde{\omega}_0 + 2 C_1^{(1)}\tilde{\alpha}_1+ 4\pi k J \\\nonumber
H_3 &=&  H_3^{(3)} \tilde{\omega}_0+2B_2^{(2)}\tilde{\alpha}_1 \\ \nonumber  
{F}_4 &=&  F_4^{(4)} + F_3^{(4)}\omega_1 + 2C_2^{(3)}\alpha_2 \\
{F}_6 &=& 
F_2^{(6)}\tilde{\omega}_4 + 2C_1^{(5)}\tilde{\alpha}_5+ 4\pi n J^3 \\ \nonumber 
H_7 &=&   
H_2^{(7)}\omega_5+ 2B_1^{(6)}\alpha_6 \\\nonumber
{F}_8 &=&  F_4^{(8)} J^2+F_3^{(8)}\omega_5 + 2C_2^{(7)}\alpha_6 \,.\nonumber
\ee
The pairs $(\alpha_r,\omega_{r-1})$ correspond to torsion cohomology with integer coefficients, 
and the pairs $(\tilde{\alpha}_r,\tilde{\omega}_{r-1})$ to torsion cohomology with local coefficients.
Note that all the coefficient fields are independent since we are not enforcing the Bianchi identities. 
Inserting the expansions into the 11d action and performing the integrals over the internal space (see Appendix \ref{HomologyAppendix}) gives the reduced 5d action
\be
S_5 &=& {1\over 2\pi} \int_{M_5} \Big[2 F_2^{(2)} (F_3^{(8)}+dC_2^{(7)}) - 2\pi k dF_4^{(8)} + 2C_1^{(1)}dF_3^{(8)}+2 H_3^{(3)} (dB_1^{(6)} + H_2^{(7)}) \nonumber \\
&+&2B_2^{(2)}dH_2^{(7)}  +2F_2^{(6)}(F_3^{(4)}-dC_2^{(3)})+2C_1^{(5)}dF_3^{(4)} +  n H_3^{(3)}F_2^{(2)} \nonumber \\
&+& k H_3^{(3)}F_2^{(6)} + \frac{2}{\pi} (H_3^{(3)}C_1^{(1)}C_1^{(5)} + B_2^{(2)}F_2^{(2)}C_1^{(5)}
+ B_2^{(2)}C_1^{(1)}F_2^{(6)})\Big].
\ee
Note that all terms not involving $k$ or $n$ are defined up to a sign,
since the integrals involving only $\omega$ and $\alpha$ are defined mod 2.
To obtain the 4d action, namely the SymTFT, we integrate out 
$F_3^{(8)}, H_2^{(7)}, F_3^{(4)}$ and $F_4^{(8)}$, which leads to the following identities for the other fields:
\be
\label{OnShellF2}
F_2^{(2)} &=&  dC_1^{(1)} \\
\label{OnShellH3}
H_3^{(3)} &=& dB_2^{(2)} \\ 
\label{OnShellF6}
F_2^{(6)} &=& dC_1^{(5)} \,.
\ee
The 5d action reduces to 
\be
S_5 &=& {1\over 2\pi} \int_{M_5} \Big[2 dC_1^{(1)}dC_2^{(7)} +2 dB_2^{(2)}dB_1^{(6)} + 2dC_2^{(3)}dC_1^{(5)} + kdB_2^{(2)}dC_1^{(5)} 
+ ndB_2^{(2)}dC_1^{(1)} \nonumber \\
&+& \frac{2}{\pi} (B_2^{(2)}dC_1^{(1)}C_1^{(5)}+B_2^{(2)}C_1^{(1)}dC_1^{(5)} + dB_2^{(2)}C_1^{(1)}C_1^{(5)})\Big]. 
\ee
The integrand is a total derivative, and the resulting 4d action on the boundary $M_4$ is given by
\be
S_4 &=& {1\over \pi} \int_{M_4} \bigg[ B_2^{(2)}d\left(B_1^{(6)} + {n\over 2}dC_1^{(1)} + {k\over 2}dC_1^{(5)}\right)
+ C_2^{(7)}dC_1^{(1)} + C_2^{(3)}dC_1^{(5)} \nonumber \\
&& \qquad \qquad \qquad \qquad \mbox{} + \frac{1}{\pi} B_2^{(2)}C_1^{(1)}C_1^{(5)} \bigg] \,.
\ee
This gives precisely the SymTFT of the orthosymplectic theory in eq. \eqref{SymTFTAction} once we make the identifications
\begin{align}
\label{symTFTfromholo}
B_1^{(6)}=\pi A_1^B,\,\,\, C_1^{(1)} = \pi A_1^{M}, \,\,\, C_1^{(5)}= \pi A_1^{C},\\ \nonumber
B_2^{(2)}=\pi A_2^B,\,\,\, C_2^{(7)} = \pi A_2^{M}, \,\,\, C_2^{(3)}= \pi A_2^{C}.
\end{align}

\section{Branes as operators}
\label{branes}

The topological surface and line operators of the 4d SymTFT correspond to branes wrapping torsion cycles of the internal space.
The branes in $AdS$ are not generally topological, but as one approaches the boundary of $AdS$ the topological sector of the
worldvolume theory is expected to dominate, and the branes become topological.
The relevant branes are the D0-brane, D2-brane, D4-brane, and D6-brane, as well as the fundamental string and the NS5-brane.
Given the transformations of the 10d gauge fields under the $\mathbb{Z}_2$ action of eq. (\ref{Z2Action}), the D0-brane, D4-brane, and fundamental string
can only wrap cycles in $H_r(\mathbb{C}P^3/{\mathbb{Z}}_2,\tilde{\mathbb{Z}})$, namely twisted cycles, and the D2-brane, D6-brane, and NS5-brane
can only wrap cycles in $H_r(\mathbb{C}P^3/\mathbb{Z}_2,\mathbb{Z})$, namely untwisted cycles.
The resulting spectrum of wrapped branes and the corresponding operators are given as 
follows:\footnote{There are additional brane wrappings that produce domain walls that source the parameters of the boundary 
theory. We mentioned these in section \ref{stringtheory}. There is also an instantonic state given by a D2-brane wrapping the torsion 3-cycle
$\Sigma_3$.}

\begin{center}
{\renewcommand{\arraystretch}{1.4}
\begin{tabular}{ |c|c|c|c|c|c|c|} 
\hline
Brane & D0 & F1 & D2 & D4 & NS5 & D6  \\ 
 \hline 
 Cycle & $\tilde{\Sigma}_0$ & $\tilde{\Sigma}_0$ & $\Sigma_1$ & $\tilde{\Sigma}_4$ & $\Sigma_5$ & $\Sigma_5$ \\
 Gauge field & $C_1^{(1)}$  & $B_2^{(2)}$  & $B_2^{(3)}$  & $C_1^{(5)}$  & $B_1^{(6)}$  & $C_2^{(7)}$   \\
 Operator & $U_M^{(1)}$ & $U_B^{(2)}$ & $\hat{U}_C^{(2)}$ & $U_C^{(1)}$ & $\hat{U}_B^{(1)}$ &
 $\hat{U}_M^{(2)}$ \\
 \hline
\end{tabular}
}
\end{center}

\begin{figure}[h!]
\centering
\includegraphics[scale=0.23, trim={0.4cm, 8cm, 3cm, 6cm}, clip]{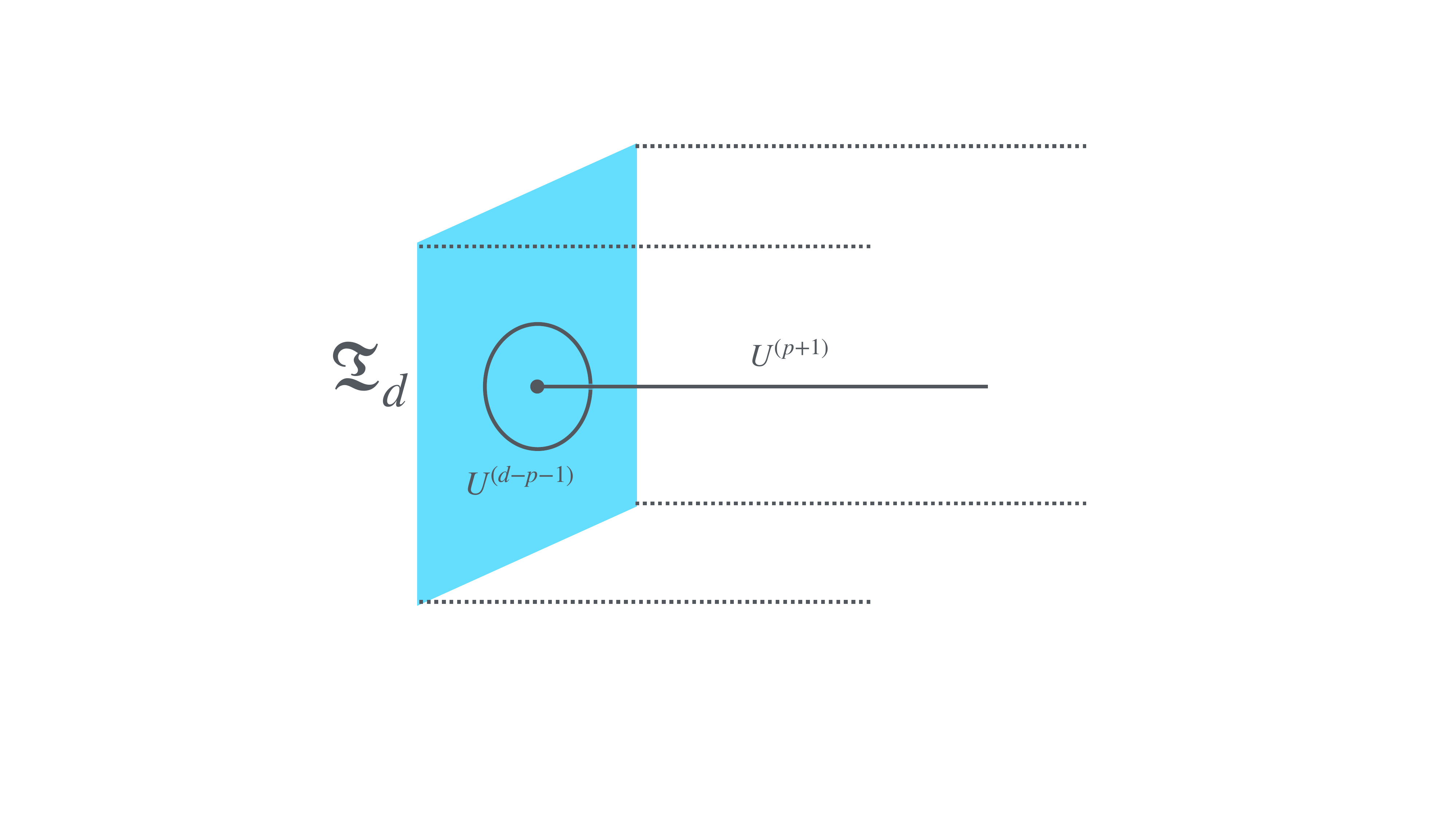}
\caption{A brane at the boundary describes a symmetry operator in the boundary theory,
and its partner brane ending at the boundary describes an operator charged under this symmetry.}
\label{chargedvstopoperators}
\end{figure}

In the 3d boundary QFT half of these branes will correspond to symmetry operators and the other half
to operators charged under the symmetries. These pair up according the $BF$ terms in the SymTFT that
determine the link-parings. In other words the pairs are $(\text{D0},\text{D6}_{\Sigma_5})$, $(\text{F1},\text{NS5}_{\Sigma_5})$,
and $(\text{D2}_{\Sigma_1},\text{D4}_{\tilde{\Sigma}_4})$.
If a given gauge field satisfies a Neumann boundary condition the corresponding wrapped brane
approaching the boundary of $AdS_4$ will describe a symmetry operator in the boundary theory.
The gauge field associated to its partner brane must then satisfy a Dirichlet boundary condition,
in which case the partner brane ending at the boundary of $AdS_4$ will describe the operator charged 
under this symmetry (see Fig. \ref{chargedvstopoperators}).\footnote{If this gauge field satisfied a Neumann BC the brane
would not really end at the bounsdary, but rather bend into the boundary. In the boundary theory this corresponds to
a {\em non-genuine} operator, which has a higher dimensional operator attached.}

For example, the $SO(2n)_{2k}\times USp(2n)_{-k}$ theory corresponds to fixing 
$\{B_2^{(2)}, C_1^{(1)}, C_1^{(5)}\}$ (namely $\{A_2^B, A_1^M, A_1^C\}$) at the boundary,
while allowing $\{B_1^{(6)}, C_2^{(7)}, C_2^{(3)}\}$ (namely $\{A_1^B, A_2^M, A_2^C\}$) to fluctuate.
In this case the NS5-brane wrapping $\Sigma_5$ and approaching the boundary corresponds to the topological line operator generating the 
1-form $\mathbb{Z}_2^{(1)}$ symmetry, and the fundamental string ending on the boundary corresponds to the Wilson line operator on which it acts.
The D6-brane wrapping $\Sigma_5$ and approaching the boundary corresponds to the topological surface operator generating $\mathbb{Z}_{2,M}^{(0)}$,
and the D0-brane ending on the boundary corresponds to the monopole operator ${\cal M}$, on which it acts.
Lastly, the D2-brane wrapping $\Sigma_1$ and approaching the boundary corresponds to the topological surface operator generating $\mathbb{Z}_{2,C}^{(0)}$,
and the D4-brane wrapping $\tilde{\Sigma}_4$ and ending on the boundary corresponds to the baryon-like operator ${\cal B}$.

\subsection{Operator TFT's from dimensional reduction of brane worldvolume theories}

To be precise, a given topological operator of the SymTFT corresponds to the path integral of the worldvolume theory of the corresponding 
wrapped brane in $AdS_4$, in the near-boundary limit.
The relevant part of D$p$-brane action in this limit is given by
\be
\label{p-braneAction}
S_{\text{Dp}} =   \int_{M_{p+1}} \left[{1\over 4\pi} e^{-\Phi} |{\cal F}_2|^2 
+ C \wedge e^{\frac{{\cal F}_2}{2\pi}}\right] ,
\ee
where ${\cal F}_2 \equiv f_2 - B_2$, and $f_2$ is the field strength of the worldvolume gauge field $a_1$.
For convenience we set $\alpha'=1$ and normalize all spacetime and worldvolume gauge fields such that their flux on a 
compact manifold is an integer multiple of $2\pi$.
We need to reduce this action for the different D$p$-branes on the appropriate cycles.
As for the bulk spacetime action, it is convenient to do this in the democratic formulation by going up one dimension \cite{Apruzzi:2023uma}.
The $(p+2)$-dimensional action for a D$p$-brane is given by
\be
S_{\text{Dp}} =   {1\over 2\pi} \int_{M_{p+2}} \left[{\cal F}_2 \wedge d{\cal F}_{p-1} - H_3\wedge {\cal F}_{p-1}
+ 2\pi {F} \wedge e^{\frac{{\cal F}_2}{2\pi}} \right] ,
\ee
where $\partial M_{p+2} = M_{p+1}$, ${F} = dC - {1\over 2\pi} H_3\wedge C$, and ${\cal F}_{p-1}$ is an independent
worldvolume $(p-1)$-form field strength. This reduces to the WZ term of the $(p+1)$-dimensional action
in eq. (\ref{p-braneAction}) after integrating out $\mathcal{F}_{p-1}$.\footnote{As in the case of the supergravity action, 
one needs to add boundary kinetic terms for the worldvolume fields 
to have a well defined variational principle. 
Here too we can neglect the boundary terms since they will only affect the kinetic terms in the reduced action, which are subdominant
when the brane approaches the boundary of $AdS$.}
We will carry out the reduction for each brane in turn.

\subsubsection{D0-brane}

The near boundary limit of the D0-brane action on the 1d manifold $M_1$ is just
\be
S_{\text{D0}} = \oint_{M_1} C_1 = \oint_{M_1} C_1^{(1)} \,.
\ee
From the holographic identification of the fields in (\ref{symTFTfromholo}) we see that
this gives the line operator $U_M^{(1)}$,
\be
U_M^{(1)}(M_1) = e^{iS_{\text{D0}}} = e^{i\oint_{M_1}C_1^{(1)}} = e^{i\pi\oint_{M_1} A_1^M} \,.
\ee

\subsubsection{D2-brane}

The next case of interest is the D2-brane wrapping the torsion 1-cycle $\Sigma_1 \sim \mathbb{R}P^1$,
and extended on a 2d submanifold $M_2 \subset AdS_4$.
The 4d action is given by
\be
S_{\text{D2}}=   {1\over 2\pi} \int_{\Sigma_1 \times M_{3}} \left[ {\cal F}_2 \wedge d{\cal F}_{1} - H_3\wedge {\cal F}_{1}
+ 2\pi {F}_4 +  F_2 \wedge {\cal F}_2 \right] ,
\ee
where $\partial M_3 = M_2$. We need to expand all fields in the basis of cohomologies of $\mathbb{R}P^1$,
which are given by $H^\ast(\mathbb{R}P^1, \mathbb{Z})=\{\mathbb{Z}, \mathbb{Z}\}$ and $H^\ast (\mathbb{R}P^1, \tilde{\mathbb{Z}})=\{0, \mathbb{Z}_2\}$. The 4-form $F_4$ is even, whereas both $F_2$ and $B_2$ are odd.
Since the 1-cycle is untwisted this implies that the worldvolume fields ${\cal F}_2$ and ${\cal F}_1$ are also odd. 
The expansion of the worldvolume fields is given by 
\be
\mathcal{F}_2 &=& \mathcal{F}_2^{(2)}\tilde{\omega}_0 + 2\eta_1 \tilde{\alpha}_1\\
\mathcal{F}_1 &=& \mathcal{F}_1^{(1)}\tilde{\omega}_0+ 2\eta_0\tilde{\alpha}_1 ,
\ee
and the expansion of the (pullbacks of the) bulk fields is given by 
\be
\label{F2PullBackExpansion}
F_2 &=&  dC_1^{(1)}\tilde{\omega}_0 + 2C_1^{(1)}\tilde{\alpha}_1\\
\label{H3PullBackExpansion}
H_3 &=& dB_2^{(2)}\tilde{\omega}_0+ 2B_2^{(2)}\tilde{\alpha}_1\\
{F}_4 &=& \left(dC_2^{(3)} + \frac{k}{2}dB_2^{(2)} + \frac{1}{\pi}B_2^{(2)}C_1^{(1)}\right)\omega_1.
\ee
Note that in pulling back the bulk fields to the D2-brane worldvolume
we impose the 10d Bianchi identities on the expansions in (\ref{expansion}).
The reduction on $\Sigma_1$ then gives
\be 
S_{\text{D2}}&=& {1\over \pi} \int_{M_3} \Big[\mathcal{F}_2^{(2)}(d\eta_0 + \mathcal{F}_1^{(1)}) + \eta_1 d\mathcal{F}_1^{(1)}
+  dC_1^{(1)}\eta_1 +  C_1^{(1)}\mathcal{F}_2^{(2)} + dB_2^{(2)}\eta_0 \nonumber\\
 && \mbox{} \qquad \qquad + B_2^{(2)}\mathcal{F}_1^{(1)} + \pi dC_2^{(3)}+ {\pi k\over 2} dB_2^{(2)} + 2 B_2^{(2)}C_1^{(1)}\Big] .
\ee
As in the case of the reduced supergravity action, each term here is defined up to a sign.
Integrating out ${\cal{F}}_1^{(1)}$ gives (up to signs)
\be
{\cal F}_2^{(2)} =  d\eta_1 + B_2^{(2)} \,.
\ee
Plugging this back in the action gives a total derivative, and we end up with an action on $M_2=\partial M_3$ given by
\be 
 S_{\text{D2}} = {1\over \pi} \oint_{M_2} \left[ \pi C_2^{(3)} + \left(\eta_0 + {\pi k\over 2}\right) B_2^{(2)} 
 + \eta_1 C_1^{(1)}  + \eta_0 d\eta_1 \right] .
\ee
This reproduces the surface operator $\hat{U}_M^{(2)}$ in (\ref{symTFToperators}) as
\be
\hat{U}_M^{(2)}(M_2) = \sum_{\eta_0,\eta_1} e^{iS^{\text{D2}}} ,
\ee
once we identify 
the worldvolume fields $\eta_{1,0}$ in terms of the cochains $\phi_{1,0}$ in the same way
as the bulk fields were related to the SymTFT cochains in eq. (\ref{symTFTfromholo}), namely 
$\eta_1 = \pi \phi_1$ and $\eta_0 = \pi \phi_0$.\footnote{Note the presence of the additional term ${\pi k\over 2} B_2^{(2)}$. 
For $k$ even this can be absorbed into a shift of $\phi_0$.
For $k$ odd, it encodes the extension of the $\mathbb{Z}_2\times \mathbb{Z}_2$ 1-form symmetry to $\mathbb{Z}_4$ for the $Spin(2n)_{2k}\times USp(2n)_{-k}$ global form
\cite{Bhardwaj:2022yxj}.
Similar terms will show up in the D6-brane and the NS5-brane theories.}

\subsubsection{D4-brane}

For the D4-brane wrapping the torsion 4-cycle $\tilde{\Sigma}_4 \sim \mathbb{R}P^4$ we have the 6d action
\be
S_{\text{D4}} =   {1\over 2\pi} \int_{\tilde{\Sigma}_4 \times M_{2}} \left[{\cal F}_2 \wedge d{\cal F}_{3} - H_3\wedge {\cal F}_{3}
+ 2\pi {F}_6 +  {F}_4 \wedge {\cal F}_2 + {1\over 4\pi} F_2 \wedge {\cal F}_2^2 \right] .
\ee
Here ${\cal F}_2$ is odd and ${\cal F}_3$ is even, so their expansions are given by\footnote{Note that the torsion
4-cycle $\tilde{\Sigma}_4$ does not itself have an integer 2-cycle.}
\begin{align*}
&\mathcal{F}_2= \mathcal{F}_2^{(2)}\tilde{\omega}_0+2\eta_1\tilde{\alpha}_1+ \mathcal{F}_0^{(2)}\tilde{\omega}_2\\&\mathcal{F}_3= \mathcal{F}_2^{(3)}\omega_1+ 2 \eta_1'\alpha_2+\mathcal{F}_0^{(3)}\omega_3.
\end{align*}
The expansions of $F_2$ and $H_3$ are the same as in eqs. (\ref{F2PullBackExpansion}) and (\ref{H3PullBackExpansion}), 
and the expansions of ${F}_4$ and ${F}_6$, imposing the 10d Bianchi identities,
are given by
\be
{F}_4 &=&  
\left(dC_2^{(3)}+ \frac{k}{2}dB_2^{(2)} + \frac{1}{\pi}B_2^{(2)}C_1^{(1)}\right)\omega_1 
+ 2C_2^{(3)}\alpha_2 \\ 
{F}_6 &=& dC_1^{(5)} \tilde{\omega}_4 \,.
\ee
The reduction on $\tilde{\Sigma}_4$ is then given by (modulo signs, as usual)
\be 
S_{\text{D4}} &=& {1\over \pi} \int_{M_2} \Big[\mathcal{F}_2^{(2)}\mathcal{F}_0^{(3)} + \eta_1 d{\cal F}_0^{(3)} 
+ {\cal F}_0^{(2)} (d\eta_1' + {\cal F}_2^{(3)}) \nonumber \\
&& \mbox{}  + B_2^{(2)} {\cal F}_0^{(3)} + \pi dC_1^{(5)} +  C_2^{(3)} {\cal F}_0^{(2)} + {1\over \pi} C_1^{(1)} \eta_1 {\cal F}_0^{(2)}
\Big] \,.
\ee
Integrating out ${\cal F}_0^{(3)}$ gives ${\cal F}_2^{(2)} = d\eta_1 + B_2^{(2)}$, and integrating out ${\cal F}_2^{(3)}$ 
gives ${\cal F}_0^{(2)}=0$. The only thing that survives is
\be
S_{\text{D4}} = \int_{M_2} dC_1^{(5)} = \oint_{M_1} C_1^{(5)} ,
\ee
where $M_1 = \partial M_2$, with no dependence on the worldvolume fields.
This reproduces the line operator $U_C^{(1)}$
\be
U_C^{(1)}(M_1) = e^{iS_{\text{D4}}} = e^{i\oint_{M_1}C_1^{(5)}} = e^{i\pi\oint_{M_1} A_1^C} \,.
\ee

\subsubsection{D6-brane}

The 8d action of the D6-brane wrapping the torsion 5-cycle $\Sigma_5$ is given by
\be
S_{\text{D6}} = \frac{1}{2\pi}\int_{\Sigma_5\times M_3} \left[\mathcal{F}_2d\mathcal{F}_5-H_3\mathcal{F}_5+ 2\pi {F}_8
+ {F}_6\wedge \mathcal{F}_2+\frac{1}{4\pi} {F}_4\wedge \mathcal{F}_2^2  
 +\frac{1}{24\pi^2} F_2\wedge \mathcal{F}_2^3 \right] . \nonumber \\
\ee
The worldvolume fields are expanded as
\be
\mathcal{F}_2 &=& \mathcal{F}_2^{(2)}\tilde{\omega}_0+2\eta_1\tilde{\alpha}_1+ \mathcal{F}_0^{(2)}\tilde{\omega}_2\\
\mathcal{F}_5 &=& \mathcal{F}_3^{(5)}\tilde{\omega}_2+ 2 \eta_2\tilde{\alpha}_3+\mathcal{F}_1^{(5)}\tilde{\omega}_4
+ 2\eta_0 \tilde{\alpha}_5 \,.
\ee
The expansions of $F_2, H_3$ and ${F}_4$ are the same as in the previous case, and the expansions of 
${F}_6$ and ${F}_8$ (satisfying the Bianchi identities) are given by
\be
{F}_6 &=&  dC_1^{(5)} \tilde{\omega}_4 + 2C_1^{(5)} \tilde{\alpha}_5 \\
{F}_8 &=& \left(dC_2^{(7)} + {n\over 2} dB_2^{(2)} + {1\over \pi} B_2^{(2)} C_1^{(5)} \right) \omega_5 \,.
\ee
The reduction on $\Sigma_5$ 
gives 
\be
S_{\text{D6}} &=& \frac{1}{\pi}\int_{M_3} \Big[
\mathcal{F}_2^{(2)}(\mathcal{F}_1^{(5)}+d\eta_0) + \eta_1 d\mathcal{F}_1^{(5)} 
+ \mathcal{F}_0^{(2)}(\mathcal{F}_3^{(5)}+d\eta_2) + dB_2^{(2)}\eta_0 + B_2^{(2)}\mathcal{F}_1^{(5)} 
 \nonumber \\
&& \mbox{} + 
\pi dC_2^{(7)} + B_2^{(2)}C_1^{(5)} +\frac{\pi n}{2}dB_2^{(2)} + dC_1^{(5)}\eta_1 + C_1^{(5)}\mathcal{F}_2^{(2)}
+  \frac{1}{\pi}C_2^{(3)}\eta_1\mathcal{F}_0^{(2)} \Big] .
\ee
Integrating out ${\cal F}_1^{(5)}$ gives ${\cal F}_2^{(2)} = d\eta_1 + B_2^{(2)}$, and integrating out ${\cal F}_3^{(5)}$
gives ${\cal F}_0^{(2)} = 0$. 
The resulting expression is again a total derivative, which gives the following action on $M_2 = \partial M_3$:
\be
S_{\text{D6}} &=& \frac{1}{\pi} \oint_{M_2} \left[\pi C_2^{(7)} + \left(\eta_0 + {\pi n\over 2}\right) B_2^{(2)} 
+ \eta_1 C_1^{(5)} + \eta_0 d\eta_1 \right] .
\ee
This reproduces the surface operator $\hat{U}_C^{(2)}$,
\be
\hat{U}_C^{(2)}(M_2) = \sum_{\eta_0,\eta_1} e^{iS_{\text{D6}}} \,.
\ee

\subsubsection{Fundamental string}

The near boundary limit of the fundamental string action on the 2d manifold $M_2$ is just
\be
S_{\text{F1}} = \oint_{M_2} B_2 = \oint_{M_1} B_2^{(2)} \,.
\ee
This gives the surface operator $U_B^{(2)}$:
\be
U_B^{(2)}(M_2) = e^{iS_{\text{F1}}} = e^{i\oint_{M_2} B_2^{(2)}} = e^{i\pi \oint_{M_2} A_2^B} \,.
\ee

\subsubsection{NS5-brane}

The remaining object is the NS5-brane wrapping the torsion 5-cycle $\Sigma_5$.
The relevant part of the NS5-brane worldvolume action is given by \cite{Bandos:2000az} 
\be
\label{NS5Action}
S_{\text{NS5}} &=& {1\over 2\pi} \int_{\Sigma_5\times M_1} \Big[
- \frac{1}{4} |\mathcal{H}_3|^2+\frac{1}{2}|\mathcal{F}_1|^2 + 2\pi B_6+\frac{1}{2} C_5C_1+  C_5\wedge dy
+\frac{1}{2}db_2C_3 \nonumber \\
&& \mbox{}  \qquad \qquad + \frac{1}{4\pi} db_2 B_2 dy+\frac{1}{4\pi} B_2 dyC_3\Big],
\ee
where $b_2$ is a worldvolume 2-form gauge field, and $y$ is a worldvolume compact scalar field,
namely a 0-form gauge field.\footnote{Strictly speaking the 3-form field strength ${\cal H}_3$ is self-dual, so its kinetic term in eq. (\ref{NS5Action}) vanishes. See \cite{Bandos:2000az} for a more careful treatment.}
The latter corresponds to the fluctuation of the NS5-brane in the M-theory direction.
The gauge invariant field strengths are given by 
\be
\mathcal{F}_1 &=& dy+C_1 \\
 \mathcal{H}_3 &=& db_2+C_3+\frac{1}{2\pi}B_2dy \,.
 \ee
We are using the same normalization conventions that we used for the D-brane actions.
We would like to find a democratic formulation of this theory in seven dimensions, as we did for the D-branes.
In particular, the equations of motion for the worldvolume fields in the 7d theory should reproduce both the equations of motion 
and the Bianchi identities for the worldvolume fields in the 6d theory.
The equation for the worldvolume scalar $y$ is given by
\be
d*{\cal F}_1 = F_6 + {1\over 2\pi} H_3 {\cal H}_3 ,
\ee
and the Bianchi identities are
\be
d{\cal H}_3 &=& F_4 + {1\over 2\pi} H_3 {\cal F}_1 \\
d{\cal F}_1 &=& F_2 \,.
\ee
The required 7d action is given by 
\be
S_{\text{NS5}} &=& {1\over 2\pi} \int_{\Sigma_5 \times M_2} \bigg[ 
{1\over 2} {\cal H}_3 \left(d{\cal H}_3   - F_4 - {1\over 2\pi} H_3 {\cal F}_1 \right) 
 - {\cal F}_5 (d{\cal F}_1  - F_2 ) \nonumber \\
&& \qquad \qquad \mbox{}   + 2\pi {H}_7 + F_6 {\cal F}_1 
- {1\over 2} \left(F_4 + {1\over 2\pi} H_3 {\cal F}_1\right) {\cal H}_3\bigg] \,,
\ee
where $\partial M_2 = M_1$.

The expansions of the RR fields and $H_3$ are as before, and the expansion of $H_7$ 
(satisfying the Bianchi identity in eq. (\ref{H7BianchiID})) is 
given by
\be
H_7 = \left(dB_1^{(6)}-\frac{n}{2}dC_1^{(1)}-\frac{k}{2}dC_1^{(5)}-\frac{1}{\pi} C_1^{(1)}C_1^{(5)}\right)\omega_5+2B_1^{(6)}\alpha_6.
\ee
The expansions of the worldvolume fields are given by 
\be
\mathcal{F}_1 &=& \mathcal{F}_1^{(1)}\tilde{\omega}_0+2\eta_0' \tilde{\alpha}_1 \\
\mathcal{F}_5 &=& 2\eta_2 \tilde{\alpha}_3+ \mathcal{F}_1^{(5)}\tilde{\omega}_4+ 2\eta_0\tilde{\alpha}_5 \\
\mathcal{H}_3 &=& {\cal H}_2^{(3)} \omega_1 + 2b_1^{(2)} \alpha_2 + {\cal H}_0^{(3)} \omega_3 \,.
\ee
The reduction to $M_2$ then gives
\be
S_{\text{NS5}} &=& {1\over \pi} \int_{M_2}\bigg[
{\cal H}_0^{(3)} \left({\cal H}_2^{(3)} + db_1^{(2)} + C_2^{(3)}  + {1\over \pi} B_2^{(2)} \eta_0'\right) 
+ b_1^{(2)} d{\cal H}_0^{(3)} \nonumber \\
&& \mbox{} + {\cal F}_1^{(5)}\left({\cal F}_1^{(1)} + d\eta_0' + C_1^{(1)}\right) 
+ {\cal F}_1^{(1)} C_1^{(5)}
+ \eta_0 \left(d{\cal F}_1^{(1)} + dC_1^{(1)}\right) + \eta_0' dC_1^{(5)} \nonumber \\
&&  \mbox{} 
+ \pi dB_1^{(6)} + {\pi n\over 2} dC_1^{(1)} + {\pi k\over 2} dC_1^{(5)} + C_1^{(1)} C_1^{(5)}
\bigg] . \nonumber \\
\ee
Integrating out ${\cal H}_2^{(3)}$ sets ${\cal H}_0^{(3)}=0$, and integrating out ${\cal F}_1^{(1)}$ sets
${\cal F}_1^{(5)} = d\eta_0' + C_1^{(5)}$. Once again this results in a total derivative, which gives
\be
S_{\text{NS5}} = {1\over \pi} \oint_{M_1} \left[\pi B_1^{(6)} + \left(\eta_0 + {\pi n\over 2}\right)C_1^{(1)} 
+ \left(\eta_0' +{\pi k\over 2}\right) C_1^{(5)} + \eta_0 d\eta_0' \right] .
\ee
This is in agreement with the line operator $\hat{U}_B^{(1)}$,
\be
\hat{U}_B^{(1)}(M_1) = \sum_{\eta_0,\eta_0'} e^{iS_{\text{NS5}}} \,.
\ee

\subsection{Non-abelian symmetry from branes}

\begin{figure}[h!]
\center
\includegraphics[height=0.3\textwidth]{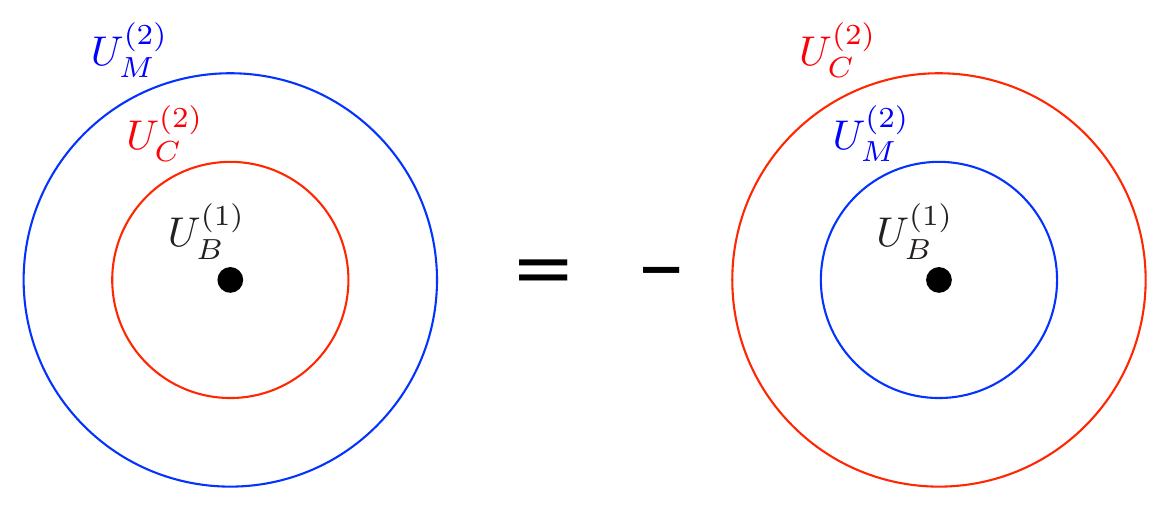}
\caption{The $M$ and $C$ surface operators commute up to a $B$-surface operator.}
\label{Commutator1}
\end{figure}

As we saw in section \ref{symTFTD8Q8}, if the set of fields fixed at the boundary is $\{A_1^B,A_1^M,A_1^C\}$,
the gauge group of the boundary theory is $(SO(2n)_{2k}\times USp(2n)_{-k})/\mathbb{Z}_2$, and the global symmetry is either $\mathbb{D}_8$ or $\mathbb{Q}_8$.
In this case the surface symmetry operators are described by F1, $\text{D2}_{\Sigma_1}$, and $\text{D6}_{\Sigma_5}$, and the corresponding local charged operators
by $\text{NS5}_{\Sigma_5}$, $\text{D4}_{\tilde{\Sigma}_4}$, and D0. The D0-brane is again dual to the monopole operator ${\cal M}$, and the D4-brane to
the baryon ${\cal B}$.
We will now see how the brane dynamics gives rise to the non-abelian global symmetry.
The crucial ingredient is that the $M$-surface operator commutes with the $C$-surface operator up to a $B$-surface operator, see eq. (\ref{SurfaceMixedFusion}).
This implies, in particular, that acting with the product of the $M$ and $C$ surface operators on the $B$-line operator, changing the order gives a minus sign, Fig.~\ref{Commutator1}.
This also shows that the local operator corresponding to the endpoint of the $B$-line transforms in the (unique) 2d representation of the global symmetry.

\begin{figure}[h!]
\centering
\includegraphics[scale=0.24, trim={1.3cm, 1.5cm, 0.3cm, 0.9cm}, clip]{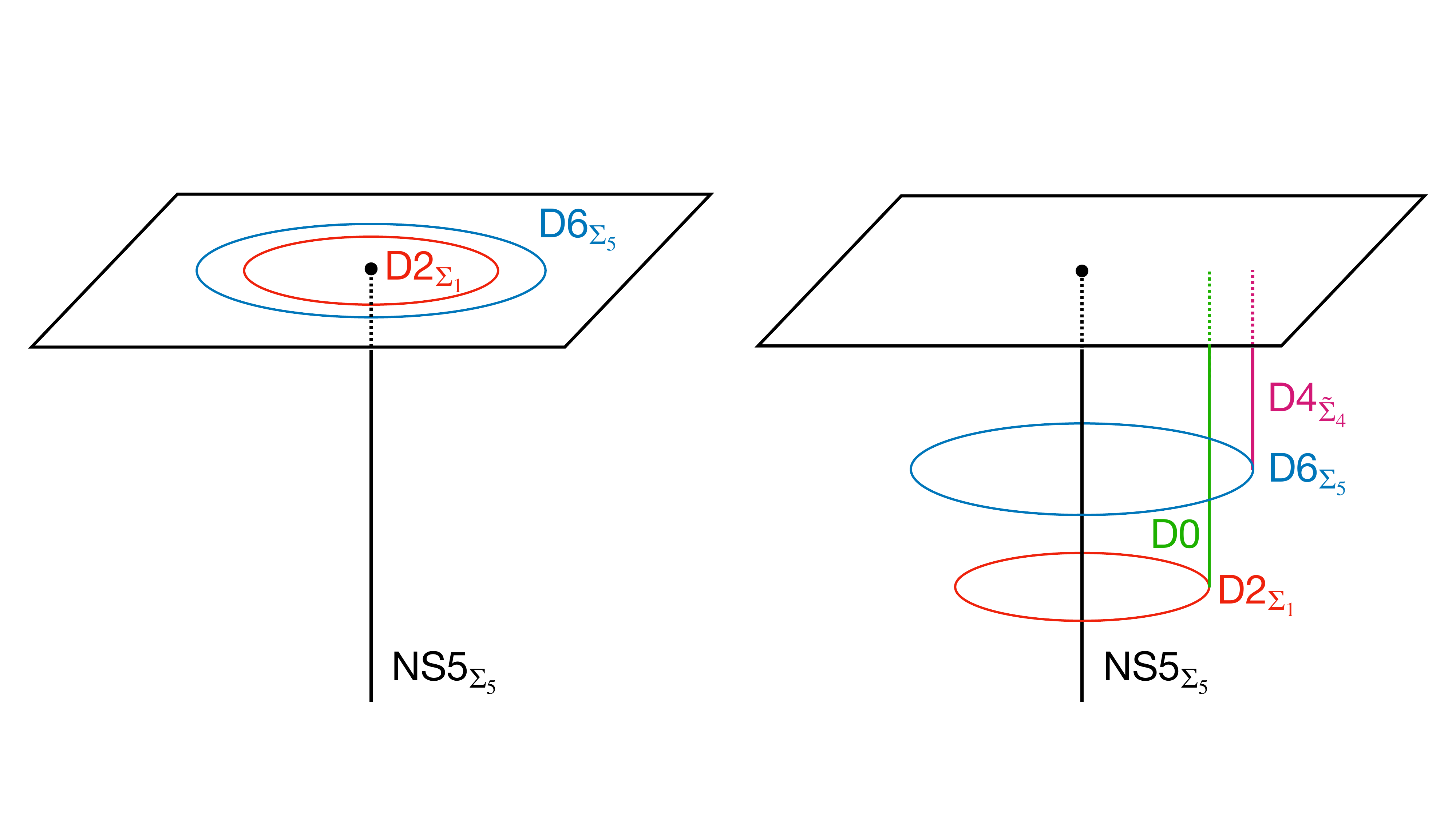}
\caption{D2$_{\Sigma_1}$ and D6$_{\Sigma_5}$ act on the NS5$_{\Sigma_5}$ before (a) and after being detached from the boundary of AdS.}\label{Commutator2}
\end{figure}

In terms of branes, $U_M^{(2)}$ is given by the D6-brane on $\Sigma_5$,
$U_C^{(2)}$ by the D2-brane on $\Sigma_1$, 
and $U_B^{(1)}$ by the NS5-brane on $\Sigma_5$, 
Fig.~\ref{Commutator2}a. 
Naively, it seems that we should be able to freely change the order of the two surface operators by moving into the bulk.
But in fact this leads to a non-trivial phase. The NS5-brane induces a magnetic tadpole in the worldvolume of both the D6-brane and the D2-brane, 
\be
 S_{\text{Dp}} \supset  \int_{\Sigma_{p-1}\times M_2}\hspace{-0.5cm} B_2 \wedge *da_1 \supset \int_{\Sigma_{p-1}\times M_2}\hspace{-0.5cm} B_2 \wedge d{a}_{p-2}
= -  \int_{\tilde{\Sigma}_{p-2}}\hspace{-0.2cm} {a}_{p-2} \quad (p=2,6) ,
\ee
where $a_{p-2}$ is the worldvolume dual magnetic potential.
The last equality follows from the fact that an NS5-brane wrapping $\Sigma_5$ and extended along the radial coordinate of $AdS_4$
sources an NSNS field strength $H_3 = \tilde{\alpha}_1 \wedge d\Omega_2$, and the relation 
$\omega_{p-1} = \tilde{\alpha}_1 \wedge \tilde{\omega}_{p-2}$.
These tadpoles must be cancelled
by a $\text{D4}_{\tilde{\Sigma}_4}$ ending on the $\text{D6}_{\Sigma_5}$, and a D0 ending on the $\text{D2}_{\Sigma_1}$, Fig.~\ref{Commutator2}b.
When we now change the order of the $\text{D6}_{\Sigma_5}$ and the $\text{D2}_{\Sigma_1}$ we pick up a sign either from the link pairing between 
the $\text{D6}_{\Sigma_5} \leftrightarrow U_M^{(2)}$ and the $\text{D0} \leftrightarrow U_M^{(1)}$ (as shown in Fig.~\ref{Commutator2}b),
or, if the original ordering was reversed, from the link pairing between the $\text{D2}_{\Sigma_1} \leftrightarrow U_C^{(2)}$ and the
$\text{D4}_{\tilde{\Sigma}_4} \leftrightarrow U_C^{(1)}$.
This sign is equivalent to the link-pairing of the $\text{NS5}_{\Sigma_5}$ with the F1.
Effectively, changing the order of the D6-brane and the D2-brane produces a fundamental string,
\be
\label{BraneCommutation}
\text{D6}_{\Sigma_5} \cdot \text{D2}_{\Sigma_1} = \text{F1} \cdot \text{D2}_{\Sigma_1} \cdot  \text{D6}_{\Sigma_5} \,.
\ee
This is the brane realization of the non-trivial commutation relation in eq. (\ref{SurfaceMixedFusion}).
Indeed, the derivation from branes mimics the derivation of the commutator in terms of the non-genuine operators $\tilde{U}_M^{(2)}, \tilde{U}_C^{(2)}$ in section \ref{SectionFusion}.
The attachment of the D4-brane to the D6-brane and the D0-brane to the D2-brane is the brane realization of the attachment of
the 3-surface operators to the 2-surface operators.

\subsection{Fractionalization from branes}

The fractionalization of order 2 elements to order 4 can also be seen from brane dynamics.
This is due to an effect first observed by Witten in \cite{Witten:1998xy}, in which a pair of $\mathbb{Z}_2$-valued branes of the same type 
annihilate, but may leave behind another type of brane.
The basic idea is that the process of annihilation of two branes wrapping a torsion cycle $\Sigma_n$ may be viewed as a single Euclidean brane wrapping
an $(n+1)$-dimensional manifold $M_{n+1}$ whose boundary consists of two copies of the cycle $\Sigma_n$, and 
there is a one-to-one map from $M_{n+1}$ to a cycle $\Sigma_{n+1}$.
If the latter cycle carries a flux this may induce a tadpole in the brane worldvolume, requiring the attachment of another type of brane.

In  \cite{Witten:1998xy}, Witten considered a pair of D5-branes wrapping an $\mathbb{R}P^4$ cycle of $\mathbb{R}P^5$ in Type IIB on $AdS_5\times \mathbb{R}P^5$.
Each of these 5-branes corresponds to a Wilson line operator in the spinor representation of the 4d dual $Spin(2n)$ SYM theory.
The annihilation process is equivalent to a 5-brane wrapping a 5d manifold with two $\mathbb{R}P^4$ boundaries, which is one-to-one with the full $\mathbb{R}P^5$.
Since there are $n$ units of 5-form flux on $\mathbb{R}P^5$, the annihilation leaves behind $n$ (mod 2) strings.
This is consistent with the fact that for even $n$ the product of two spinors is in the identity class, and for odd $n$ it is in the vector class.

In our case the relevant branes are the D2-brane wrapping the torsion 1-cycle $\Sigma_1$ and the D6-brane wrapping the torsion 5-cycle $\Sigma_5$.
Two D2-branes wrapping $\Sigma_1$ will leave behind $k$ (mod 2) fundamental strings, due to the $k$ units of $F_2$ flux on the integral 2-cycle
$\mathbb{C}P^1 \subset \mathbb{C}P^3/\mathbb{Z}_2$, and two D6-branes wrapping $\Sigma_5$ will leave behind $n$ (mod 2) fundamental strings, 
due to the $n$ units of $F_6$ flux on $\mathbb{C}P^3/\mathbb{Z}_2$. Symbolically,
\begin{align}
\label{BraneSelfFusion}
&\text{F1} \cdot \text{F1} = 1 \nonumber \\
&\text{D6}_{\Sigma_5} \cdot  \text{D6}_{\Sigma_5}  = n \; (\text{mod 2}) \; \text{F1} \\
&\text{D2}_{\Sigma_1} \cdot  \text{D2}_{\Sigma_1}  = k \; (\text{mod 2}) \; \text{F1} \,. \nonumber
\end{align}
These reproduce precisely the self-fusions of the surface operators $U_B^{(2)}, \tilde{U}_M^{(2)}, \tilde{U}_C^{(2)}$ in eq. (\ref{SurfaceSelfFusion}).
Combining the last two relations in eq. (\ref{BraneSelfFusion}) with the commutation relation in eq. (\ref{BraneCommutation}) we also get
\be
(\text{D6}_{\Sigma_5} \cdot \text{D2}_{\Sigma_1})^2 = (n+k+1) \; (\text{mod 2}) \; \text{F1} \,,
\ee
in agreement with eq. (\ref{UMCsquared}). In other words, if either $n$ or $k$ are even there is only one order 4 element, and the group is $\mathbb{D}_8$,
and if both $n$ and $k$ are odd there are three order 4 elements, and the group is $\mathbb{Q}_8$.

\subsection{Representations from branes}

Both $\mathbb{D}_8$ and $\mathbb{Q}_8$ have exactly four one-dimensional irreducible representations (including the trivial representation),
and a single two-dimensional irreducible representation (see Appendix \ref{Groups}).
The local operators corresponding to the D0, $\text{D4}_{\tilde{\Sigma}_4}$ and $\text{NS5}_{\Sigma_5}$ should fit into these representations.
The D0 and the $\text{D4}_{\tilde{\Sigma}_4}$ correspond to two of the non-trivial 1d representations. This is clear from the fact that they are acted on non-trivially 
by only one of the symmetry branes. The D0 links non-trivially only with the $\text{D6}_{\Sigma_5}$, and the $\text{D4}_{\tilde{\Sigma}_4}$
links non-trivially only with the $\text{D2}_{\Sigma_1}$. The third non-trivial 1d representation corresponds to the bound state $\text{D0} \text{D4}_{\tilde{\Sigma}_4}$.
This is also consistent with the products of the 1d representations.

The $\text{NS5}_{\Sigma_5}$ corresponds to the 2d representation.
We already encountered evidence for this from the fact that the action of the $\text{D6}_{\Sigma_5}$ and the $\text{D2}_{\Sigma_1}$ on the 
$\text{NS5}_{\Sigma_5}$ was non-commutative. This is only possible in a higher-dimensional representation.
Another hint comes from the annihilation effect.
A pair of  $\text{NS5}_{\Sigma_5}$ branes will leave behind $n$ (mod 2) D0-branes, due to the flux on $\mathbb{C}P^3/\mathbb{Z}_2$,
as well as $k$ (mod 2) D4-branes wrapping $\tilde{\Sigma}_4$, due to the flux on $\mathbb{C}P^1 \subset \mathbb{C}P^3/\mathbb{Z}_2$,
\be
\text{NS5}_{\Sigma_5} \cdot \text{NS5}_{\Sigma_5} = n  \text{D0} + k  \text{D4}_{\tilde{\Sigma}_4} \,.
\ee
This reproduces the self-fusion of the line operator $\tilde{U}_B^{(1)}$ in eq. (\ref{LineSelfFusion}).
As we noted before, this is not quite the correct relation between the representations.
For that, we needed to consider the non-invertible version of the line operator, $\hat{U}_B^{(1)}$ in eq. (\ref{RepDecomposition}).


\section{Conclusions and outlook}
\label{concl}

In this paper we studied the global symmetries and symmetry theory of the 3d $\mathcal{N}=5$ orthosymplectic 
Chern-Simons theories. This class of theories provides another interesting example in which the symmetry theory
can be derived from the holographic dual string theory description. 
We showed that the 4d SymTFT action is recovered by reducing 10d Type IIA supergravity on 
the internal space $\mathbb{C}P^3/\mathbb{Z}_2$, and identified all the symmetry operators with  
branes wrapping torsion cycles.
We also established the correspondence between the different global structures of the orthosymplectic gauge group
and the consistent boundary conditions for the bulk gauge fields.
Of particular interest was the 3d theory in which the global structure of the gauge group is $(SO(2n)_{2k}\times USp(2n)_{-k})/\mathbb{Z}_2$, which exhibits a non-abelian global symmetry given by the dihedral group $\mathbb{D}_8$, if either $n$ or $k$ are even, or by the quaternion group $\mathbb{Q}_8$, if both $n$ and $k$ are odd.
We demonstrated how the structure of these groups arises in the string theory description from brane dynamics.

There are a number of generalizations that would be interesting to explore.
The most obvious ones are to more general orthosymplectic theories with unequal ranks and unequal Chern-Simons
levels, as well as to theories with an $\mathfrak{so}(2n+1)$ factor.
In the dual string theory, this requires turning on the additional fluxes discussed in the beginning of section \ref{stringtheory}.
The duality in this more general setting was understood in \cite{Aharony:2008gk}, but it would be interesting to study
how the global symmetry structure emerges in the holographic picture.
An even more interesting case involves M2-branes on $\mathbb{C}^4/\mathbb{E}_k$ (with $k=6,7$ or $8$),
which also preserves ${\cal N}=5$ supersymmetry \cite{Morrison:1998cs}.
For large $N$ this would imply a duality between
M-theory on $AdS_4\times S^7/\mathbb{E}_k$ and new 3d ${\cal N}=5$ SCFTs.
We do not (yet) have a Lagrangian description for these theories, and therefore do not a-priori know their 
global symmetry structures.
Perhaps one can use holography to study this.
A first step might be to understand the Type IIA dual, so that we could use the tools developed in this paper.

\subsection*{Acknowledgements}

We thank Jonathan J. Heckman, Max H\"ubner, Shlomo Razamat, Xingyang Yu, Gabi Zafrir and Yunqin Zheng for discussions. This work is supported in part by the Israel Science Foundation under grant No. 1254/22,
and by the US-Israel Binational Science Foundation under grant No. 2022100. F.M is partially supported by the Spanish
national grant MCIU-22-PID2021-123021NB-I00
\addcontentsline{toc}{section}{Bibliography}

\appendix

\section{Properties of the groups $\mathbb{D}_8$ and $\mathbb{Q}_8$}
\label{Groups}

The dihedral group $\mathbb{D}_8$ is the group of eight elements describing the symmetries of the square, which are 
generated by a 90 degree rotation and a reflection.
The quaternion group $\mathbb{Q}_8$ is the group of eight elements $\{1,-1,i,-i,j,-j,k,-k\}$,
where $i^2=j^2=k^2=-1$, $ij=-ji=k$, $jk=-kj=i$, and $ki=-ik=j$.
In standard notation, the generators of either group are denoted by $a$ and $x$,
and the two groups are defined as 
\be
\mathbb{D}_8 &=& \{ a,x: a^4=1, \, x^2=1, \, xax^{-1}= a^3 \} \\
\mathbb{Q}_8 &=& \{a,x: a^4=x^4=1, \, a^2=x^2, \, xax^{-1}=a^3\} \,.
\ee
Other than the trivial representation, both groups have three 1-dimensional irreducible representations and one 2-dimensional
irreducible representation. Table~\ref{RepTable} shows the representations explicitly.
A product of two 1-dimensional representations gives another 1-dimensional representation.
\begin{table}[h!]
\centering
\begin{tabular}{|c|c|c|c|c|c|c|} 
\hline
 & trivial & $\langle a \rangle$-kernel &  $\langle a^2,x \rangle$-kernel & $\langle a^2,ax \rangle$-kernel & $\mathbb{D}_8$ 2d & $\mathbb{Q}_8$ 2d \\ 
 \hline
 1 & 1 & 1 & 1 & 1 & $\left( \begin{array}{cc} 1 & 0 \\ 0 & 1 \end{array} \right)$ & $\left( \begin{array}{cc} 1 & 0 \\ 0 & 1 \end{array} \right)$\\
  \hline
 $a$ & $1$ & $1$ & $-1$ & $-1$ & $\left( \begin{array}{cc} 0 & 1 \\ -1 & 0 \end{array} \right)$ & $\left( \begin{array}{cc} 0 & 1 \\ -1 & 0 \end{array} \right)$ \\
  \hline
 $a^2$ & $1$ & $1$ & $1$ & $1$ & $\left( \begin{array}{cc} -1 & 0 \\ 0 & -1 \end{array} \right)$ & $\left( \begin{array}{cc} -1 & 0 \\ 0 & -1 \end{array} \right)$\\
  \hline
  $a^3$ & $1$ & $1$ & $-1$ & $-1$ & $\left( \begin{array}{cc} 0 & -1 \\ 1 & 0 \end{array} \right)$ & $\left( \begin{array}{cc} 0 & -1 \\ 1 & 0 \end{array} \right)$\\
  \hline
 $x$ & $1$ & $-1$ & $1$ & $-1$ & $\left( \begin{array}{cc} 1 & 0 \\ 0 & -1 \end{array} \right)$ & $\left( \begin{array}{cc} -i & 0 \\ 0 & i \end{array} \right)$\\
 \hline
 $ax$ & $1$ & $-1$ & $-1$ & $1$ & $\left( \begin{array}{cc} 0 & -1 \\ -1 & 0 \end{array} \right)$ & $\left( \begin{array}{cc} 0 & i \\ i& 0 \end{array} \right)$\\
 \hline
 $a^2x$ & $1$ & $-1$ & $1$ & $-1$ & $\left( \begin{array}{cc} -1 & 0 \\ 0 & 1 \end{array} \right)$ & $\left( \begin{array}{cc} i & 0 \\ 0 & -i \end{array} \right)$\\
 \hline
 $a^3x$ & $1$ & $-1$ & $-1$ & $1$ & $\left( \begin{array}{cc} 0 & 1 \\ 1 & 0 \end{array} \right)$ & $\left( \begin{array}{cc} 0 & -i \\ -i& 0 \end{array} \right)$\\
 \hline
\end{tabular}
\caption{Representation table of $\mathbb{D}_8$ and $\mathbb{Q}_8$}\label{Table2}
\label{RepTable}
\end{table}
A product of any 1-dimensional representation with the 2-dimensional representation gives back the same 2-dimensional representation.
The product of the 2-dimensional representation with itself gives the direct sum of the four 1-dimensional representations,

\be
\label{2dtimes2d}
\langle\text{2d}\rangle \times \langle\text{2d}\rangle = \langle 1 \rangle + \langle a \rangle +  \langle a^2, x \rangle +  \langle a^2, ax \rangle \,.
\ee

\section{Homology and cohomology of $\mathbb{C}P^3/\mathbb{Z}_2$}
\label{HomologyAppendix}

We can infer the homology groups of $\mathbb{C}P^3/\mathbb{Z}_2$ starting from $\mathbb{C}P^3$ and studying the effect of the  
$\mathbb{Z}_2$ projection.
Let $(z_1,z_2,z_3,z_4)$ be the homogenous complex coordinates on $\mathbb{C}P^3$, namely we identify
\be
\label{CP3}
(z_1,z_2,z_3,z_4) \sim \lambda (z_1,z_2,z_3,z_4) \; \mbox{for all} \; \lambda\in \mathbb{C}^\ast \,.
\ee
The non-trivial homology groups of $\mathbb{C}P^3$ are all even dimensional 
and generated by $\mathbb{C}P^k\subset \mathbb{C}P^3$:
\begin{center}
\begin{tabular}{ |c|c|c|c|c|c|c|c|} 
\hline
$r$ & 0 & 1 & 2 & 3 & 4 & 5 & 6 \\ 
\hline
$H_r(\mathbb{C}P^3,\mathbb{Z})$ & $\mathbb{Z}$ & 0 & $\mathbb{Z}$ & 0 &  $\mathbb{Z}$ & 0  & $\mathbb{Z}$ \\ 
 \hline
\end{tabular}
\end{center}
The dual cohomology groups are generated by $J^k$, where $J$ is the K\"ahler 2-form of $\mathbb{C}P^3$.
For example, the basic 2-cycle is given by a $\mathbb{C}P^1\subset \mathbb{C}P^3$ and can be parameterized by any two of the $z_i$, setting the other two to zero, e.g.  $(z_1,z_2,0,0)$ or $(z_1,0,z_3,0)$. All the different parameterizations are related by the 
$SU(4)$ isometry of $\mathbb{C}P^3$.

The quotient space $\mathbb{C}P^3/\mathbb{Z}_2$ is defined by the additional identification
\be
\label{Z2transformation}
(z_1,z_2,z_3,z_4) \sim (iz_2^*,-iz_1^*,iz_4^*,-iz_3^*) \,.
\ee
This maps $J \rightarrow -J$ and breaks the isometry group to $USp(4) \subset SU(4)$.
The 0-cycle and 4-cycle survive the projection, but 
the 2-cycle and 6-cycle are removed from the homology with integer coefficients.
On the other hand they appear in the corresponding homology groups with local coefficients, 
$H_{2,6}(\mathbb{C}P^3/\mathbb{Z}_2,\tilde{\mathbb{Z}})$.

There are also new torsion-valued cycles.
In the odd-dimensional case these appear in homology with integer coefficients as follows.
The generator of $H_1(\mathbb{C}P^3/\mathbb{Z}_2,\mathbb{Z})$ is an $\mathbb{R}P^1$ parametrized by 
$(x_1,iy_2,0,0)$ with $x_1,y_1\in \mathbb{R}$.
The generator of $H_3(\mathbb{C}P^3/\mathbb{Z}_2,\mathbb{Z})$ is an $\mathbb{R}P^3$ parametrized by 
$(x_1,iy_2,x_3,iy_2)$ with $x_i,y_i\in\mathbb{R}$.
The generator of $H_5(\mathbb{C}P^3/\mathbb{Z}_2,\mathbb{Z})$ is an
$\mathbb{R}P^5$ parametrized by $(x_1,iy_2,z_3,z_4)$ with $x_1,y_2\in\mathbb{R}$ and $z_3,z_4\in\mathbb{C}$. Of course, there are equivalent parameterizations of these cycles that are related to the ones above
by an $USp(4)$ isometry transformation.
Moreover, these are all $\mathbb{Z}_2$ valued since two copies of each one can be viewed as the same subspace 
of the original $\mathbb{C}P^3$, which has trivial odd-dimensional homology groups.

In the even-dimensional case the torsion-valued cycles appear in homology with local coefficients.
We can give an explicit construction for the torsion 2-cycle in $H_2(\mathbb{C}P^3/\mathbb{Z}_2,\tilde{\mathbb{Z}})$ as follows.
Recall that the covering space $\mathbb{C}P^3$ has a 2-cycle $\mathbb{C}P^1 \subset \mathbb{CP}^3$ that can be parameterized 
by any pair of the four homogeneous coordinates $z_i$ of $\mathbb{C}P^3$. This is unique since all
parameterizations are related by the $SU(4)$ isometry of $\mathbb{C}P^3$.
On the other hand, the map in eq. (\ref{Z2transformation}) defining the quotient $\mathbb{C}P^3/\mathbb{Z}_2$ distinguishes the 2-cycles
parameterized by $(z_1,z_2)$ or $(z_3,z_4)$, from the 2-cycles parameterized by other pairs, like $(z_1,z_3)$.
The former 2-cycles map to themselves, leading to a reduced 2-cycle $\mathbb{C}P^1/\mathbb{Z}_2 \approx \mathbb{R}P^2$,
whereas the latter maps to a different parametrization, maintaining the geometry $\mathbb{C}P^1$.
Crucially, the two types of cycles are not related by the reduced isometry $USp(4)$ of $\mathbb{C}P^3/\mathbb{Z}_2$.
We can see that the $\mathbb{R}P^2$ cycle is $\mathbb{Z}_2$-valued as follows.
Define a standard set of angular coordinates on $\mathbb{C}P^3$ as
\be
z_1 &=& \cos\xi \cos{\theta_1\over 2} e^{i(\psi+\phi_1)/2} \nonumber\\
z_2 &=& \cos\xi \sin{\theta_1\over 2} e^{i(\psi-\phi_1)/2} \nonumber\\
z_3 &=& \sin\xi \cos{\theta_2\over 2} e^{i(-\psi+\phi_2)/2} \nonumber\\
z_4 &=& \sin\xi \sin{\theta_2\over 2} e^{i(-\psi-\phi_1)/2}
\ee
where $0\leq\xi\leq \pi/2$, $0\leq \theta_i \leq \pi$, $0\leq \psi, \phi_i < 2\pi$. 
The $\mathbb{C}P^1$ cycle parameterized by $(z_1,z_2)$ corresponds to $\xi=0$, and the one parameterized by $(z_3,z_4)$
corresponds to $\xi=\pi/2$. 
Now consider the 3d subspace $M_3 \subset \mathbb{C}P^3$ defined by $\psi = 0$, $\theta_1=\theta_2=\theta$,
and $\phi_1=\phi_2=\phi$.
The boundary of $M_3$ is given by the union of the first $\mathbb{C}P^1$ and the orientation reversal of the second $\mathbb{C}P^1$,
\be
\partial M_3 = \mathbb{C}P^1 \cup \overline{\mathbb{C}P^1} \,.
\ee
This of course is consistent with the fact that $H_2(\mathbb{C}P^3,\mathbb{Z})=\mathbb{Z}$, since the RHS is trivial in homology.
The map (\ref{Z2transformation}) takes $\psi \rightarrow  - \psi$, $\theta_i \rightarrow \pi - \theta_i$, 
and $\phi_i \rightarrow \phi_i + \pi$, and therefore maps $M_3$ to itself, 
producing the quotient space $M_3/\mathbb{Z}_2$. The boundary of this space is given by
\be
\partial (M_3/\mathbb{Z}_2) = \mathbb{C}P^1/\mathbb{Z}_2 \cup \mathbb{C}P^1/\mathbb{Z}_2 \,,
\ee
which implies that the homology class of $\mathbb{C}P^1/\mathbb{Z}_2$ is a mod 2 element.
Note that $\mathbb{C}P^1/\mathbb{Z}_2\approx \mathbb{R}P^2$ is a non-orientable manifold.
One should contrast this with the $\mathbb{C}P^1$
that generates the free part of $H_2(\mathbb{C}P^3/\mathbb{Z}_2, {\mathbb{Z}})$, which can be parameterized by the other possible pairs: $(z_1,z_3), (z_1,z_4), (z_2,z_3)$, or $(z_2,z_4)$.
These are not quotiented by (\ref{Z2transformation}) since they map pairwise to each other, parametrizing a $\mathbb{C}P^1\subset \mathbb{C}P^3/\mathbb{Z}_2$ which is not related to the cycle parameterized by $(z_1,z_2)$ (or $(z_3,z_4)$) by $USp(4)$.
The second homology group with local coefficients is therefore given by 
$H_2(\mathbb{C}P^3/\mathbb{Z}_2, \tilde{\mathbb{Z}}) = \mathbb{Z} \oplus \mathbb{Z}_2$.

The remaining even-dimensional tosion-valued cycles in twisted-homology can be inferred using known properties 
of homology and cohomology groups with integer or local coefficients.
First, for a non-orientable closed and connected $n$-dimensional manifold $M$ (like $\mathbb{C}P^3/\mathbb{Z}_2$) we have that
\cite{Hatcher}:
\be
\label{HomologyID1}
H_r(M,\tilde{\mathbb{Z}}) & \approx  & H^{n-r}(M,\mathbb{Z}) \,. 
\ee
Second, for any chain complex $M$ with finitely generated homology groups we have that
\be
\label{HomologyID2}
H^r(M,\mathbb{Z}) & \approx & \left(H_r(M,\mathbb{Z})/\mbox{Tor} H_r(M,\mathbb{Z})\right) \oplus \mbox{Tor} H_{r-1}(M,\mathbb{Z}) \,,
\ee
and therefore
\be
\label{HomologyID3}
  \mbox{Tor} H_{r-1}(M,\mathbb{Z}) \approx    \mbox{Tor} H^r(M,\mathbb{Z}) \approx  \mbox{Tor} H_{n-r}(M,\tilde{\mathbb{Z}})\,.
\ee
For $M=\mathbb{C}P^3/\mathbb{Z}_2$ we therefore find that
\be
\mbox{Tor} H_{2k}(\mathbb{C}P^3/\mathbb{Z}_2,\tilde{\mathbb{Z}}) 
& \approx & \mbox{Tor} H_{5-2k}(\mathbb{C}P^3/\mathbb{Z}_2,{\mathbb{Z}}) \nonumber \\
& = & \mathbb{Z}_2 \,.
\ee
Using the same kind of reasoning one can show that the odd homologies with local coefficients are all torsion free, and therefore trivial.
The homology groups of $\mathbb{C}P^3/\mathbb{Z}_2$ are summarized in Table~\ref{AppendixHomologies}.
For completeness we also list the cohomology groups.
This table is reproduced in section 3.
\begin{table}
\centering
{\renewcommand{\arraystretch}{1.2}
\begin{tabular}{ |c|c|c|c|c|c|c|c|} 
\hline
$p$ & 0 & 1 & 2 & 3 & 4 & 5 & 6 \\ 
 \hline
$H_p(\mathbb{C}P^3/\mathbb{Z}_2,\mathbb{Z})$ & $\mathbb{Z}$ & $\mathbb{Z}_2$ & 0 & $\mathbb{Z}_2$ &  $\mathbb{Z}$ & $\mathbb{Z}_2$  & 0 \\ 
$H^p(\mathbb{C}P^3/\mathbb{Z}_2,\mathbb{Z})$ & $\mathbb{Z}$ & 0  & $\mathbb{Z}_2$ & 0 &  $\mathbb{Z}\oplus \mathbb{Z}_2$ & 0  & $\mathbb{Z}_2$ \\ 
\hline 
$H_p(\mathbb{C}P^3/\mathbb{Z}_2,\tilde{\mathbb{Z}})$ & $\mathbb{Z}_2$ & 0 & $\mathbb{Z} \oplus \mathbb{Z}_2$ &  0 & $\mathbb{Z}_2$ & 0  & $\mathbb{Z}$ \\ 
$H^p(\mathbb{C}P^3/\mathbb{Z}_2,\tilde{\mathbb{Z}})$ & 0 & $\mathbb{Z}_2$ & $\mathbb{Z}$ &  $\mathbb{Z}_2$ & 0 & $\mathbb{Z}_2$  & $\mathbb{Z}$\\
 \hline
\end{tabular}
}
\caption{Homology and cohomology groups of $\mathbb{C}P^3/\mathbb{Z}_2$.}\label{homologiesCP3Z2}
\label{AppendixHomologies}
\end{table}

As another consistency check one can use the long exact homology sequence for non-orientable manifolds:
\be
\cdots \rightarrow H_k(X,\tilde{\mathbb{Z}}) \rightarrow H_k(X',\mathbb{Z}) \rightarrow H_k(X,\mathbb{Z}) \rightarrow H_{k-1}(X,\tilde{\mathbb{Z}})
\rightarrow \cdots
\ee
where $X$ is a non-orientable manifold, and $X'$ is its orientable double-cover.
In our case $X=\mathbb{C}P^3/\mathbb{Z}_2$ and $X'=\mathbb{C}P^3$.
For example a consistency check for $H_2(\mathbb{C}P^3/\mathbb{Z}_2,\tilde{\mathbb{Z}})$ is provided by a piece of this sequence:
\be
\begin{array}{ccccccccc}
H_3(X',\mathbb{Z}) &\rightarrow & H_3(X,\mathbb{Z}) & \rightarrow & H_2(X,\tilde{\mathbb{Z}}) & \rightarrow & H_2(X',\mathbb{Z}) 
& \rightarrow & H_2(X,\mathbb{Z}) \\
\parallel &  & \parallel & & \parallel & & \parallel & & \parallel \\
0 & \rightarrow & \mathbb{Z}_2 & \rightarrow & \mathbb{Z} \oplus \mathbb{Z}_2 & \rightarrow & \mathbb{Z} & \rightarrow & 0
\end{array}
\ee
The second map is the identity map from $\mathbb{Z}_2$ to $\mathbb{Z}_2$, and the second map is the identity map
from $\mathbb{Z}$ to $\mathbb{Z}$.

\subsection{Integrals}
\label{Integrals}
The volume form of $\mathbb{C}P^3/\mathbb{Z}_2$ coincides with $J^3$ and it is normalized to integrate to $\frac{1}{2}$
\be
\int_{\mathbb{C}P^3/\mathbb{Z}_2} J^3 = {1\over 2}. 
\ee
Torsional classes pair on $\mathbb{C}P^3/\mathbb{Z}_2$ through their corresponding holonomies  \cite{Camara:2011jg, Hsieh:2020jpj} 
\be
\int_{\mathbb{C}P^3/\mathbb{Z}_2} \tilde{\alpha}_{2p+1} \wedge \omega_{6-2p-1} = 
\int_{\mathbb{C}P^3/\mathbb{Z}_2} {\alpha}_{2p} \wedge \tilde{\omega}_{6-2p} 
=1\,\, \text{mod}\,\, 2.
\ee
In particular, as in the $\mathbb{R}P^5$ case \cite{10.1215/kjm/1250517912, GarciaEtxebarria:2022vzq}, the generator of $\text{Tor}H^{2n+1}(\mathbb{C}P^3/\mathbb{Z}_2,\tilde{ \mathbb{Z}})$ (resp. $\text{Tor}H^{2n}(\mathbb{C}P^3/\mathbb{Z}_2, \mathbb{Z})$) is the $(2n+1)-$th (resp. $2n-$th) product of $\tilde{\alpha}_1\in H^1(\mathbb{C}P^3/\mathbb{Z}_2, \tilde{\mathbb{Z}})$. As a consequence, the holonomies $\tilde{\omega}_{2n}$ (resp. $\omega_{2n-1}$) associated with the $\tilde{\alpha}_{2n+1}$ (resp. $\alpha_{2n}$) can be equivalently described as the product $(-1)^{n-1}\tilde{\omega}_0\tilde{\alpha}_1^{2n}$ (resp. $(-1)^{n-1}\tilde{\omega}_0\tilde{\alpha}_1^{2n-1}$).\\
In addition, the following integrals are non-zero and normalized to $\frac{1}{2}$

\be\label{specialintegrals}
\int_{\mathbb{C}P^3/\mathbb{Z}_2} \omega_p\wedge \omega_{2r-p} \wedge J^{3-r}=\frac{1}{2}, \,\,\,\,\,\, \int_{\mathbb{C}P^3/\mathbb{Z}_2} J\wedge J \wedge \tilde{\omega}_2 = 
\frac{1}{2}. 
\ee


\bibliography{ArXiv_v2bib}
\bibliographystyle{JHEP} 
\end{document}